\documentclass[12pt]{article}

\usepackage{scicite}
\usepackage{times}
\usepackage{booktabs}
\usepackage{bbold}
\usepackage{graphicx}
\usepackage{amsmath,amssymb}
\usepackage{dsfont}
\usepackage{color}
\usepackage[normalem]{ulem}

\newenvironment{sciabstract}{%
\begin{quote} \bf}
{\end{quote}}

\newcommand{\ket}[1]{ \left |#1 \right\rangle}

\newcounter{lastnote}

\title{Critical thermalization of a disordered dipolar spin system in diamond} 

\author
{G. Kucsko,$^{1\dagger}$ S. Choi,$^{1\dagger}$ J. Choi,$^{1,2\dagger}$ P. C. Maurer,$^{3}$ H. Zhou,$^{1}$ R. Landig,$^{1}$ \\
H. Sumiya,$^{4}$ S. Onoda,$^{5}$ J. Isoya,$^{6}$ F. Jelezko,$^{7}$ E. Demler,$^{1}$ N. Y. Yao,$^{8}$ M. D. Lukin$^{1\ast}$\\
\\
\normalsize{$^{1}$Department of Physics, Harvard University,}\\
\normalsize{Cambridge, MA 02138, USA}\\
\normalsize{$^{2}$School of Engineering and Applied Sciences, Harvard University,}\\
\normalsize{Cambridge, MA 02138, USA}\\
\normalsize{$^{3}$Department of Physics, Stanford University,}\\
\normalsize{Stanford, California 94305, USA}\\
\normalsize{$^{4}$Sumitomo Electric Industries Ltd., Itami, Hyougo, 664-0016, Japan}\\
\normalsize{$^{5}$Takasaki Advanced Radiation Research Institute,}\\ \normalsize{National Institutes for Quantum and Radiological Science and Technology,}\\
\normalsize{1233 Watanuki, Takasaki, Gunma 370-1292, Japan}\\
\normalsize{$^{6}$Research Centre for Knowledge Communities}\\ \normalsize{University of Tsukuba, Tsukuba, Ibaraki 305-8550, Japan}\\
\normalsize{$^{7}$Institut f{\"u}r Quantenoptik, Universit{\"a}t Ulm,}\\
\normalsize{89081 Ulm, Germany}\\
\normalsize{$^{8}$Department of Physics, University of California Berkeley,}\\
\normalsize{Berkeley, California 94720, USA}\\
\\
\normalsize{$^\dagger$These authors contributed equally to this work.}
\\
\normalsize{$^\ast$To whom correspondence should be addressed}
\\
\normalsize{E-mail: lukin@physics.harvard.edu.}
}

\date{}

\begin{document} 

\sloppy

\baselineskip24pt

\maketitle 

\begin{sciabstract}
Statistical mechanics underlies our understanding of macroscopic quantum systems. It is based on the assumption that out-of-equilibrium systems rapidly approach their equilibrium states, forgetting any information about their microscopic initial conditions. 
This fundamental paradigm is challenged by disordered systems, in which a slowdown or even absence of thermalization is expected. 
We report the observation of critical thermalization in a three dimensional ensemble of $\sim10^6$ electronic spins coupled via dipolar interactions. 
By controlling the spin states of nitrogen vacancy color centers in diamond, we observe slow, sub-exponential relaxation dynamics and identify a regime of power-law decay with disorder-dependent exponents; this behavior is modified at late times owing to many-body interactions. These observations are quantitatively explained by a resonance counting theory that incorporates the effects of both disorder and interactions.
\end{sciabstract}

Nearly six decades ago, Anderson predicted that the interplay between long-range couplings and disorder in quantum systems can lead to a novel regime of slow, sub-diffusive thermalization~\cite{anderson1958absence}. 
This is in stark contrast  to both conventional ergodic systems and  disordered systems with short-range hopping, where disorder can arrest dynamics, resulting in the breakdown of ergodicity.
Termed Anderson localization, the latter effect has been observed in systems ranging from acoustic and optical waves to cold atomic gases~\cite{schwartz2007transport,billy2008direct,roati2008anderson}; more recently, it has been shown that localization can  persist even in strongly-interacting, isolated quantum systems, a phenomenon dubbed many-body localization~\cite{nandkishore2015many,
schreiber2015observation,smith2016many}. In addition to raising fundamental questions, such systems have also become a basis for the exploration of novel non-equilibrium phases of matter, including Floquet symmetry protected topological phases~\cite{potter2016classification} and discrete time crystals~\cite{Zhang:2016uw,choi2016observation}. 

The addition of long-range couplings tends to facilitate delocalization, leading to a regime where ergodicity and localization compete~\cite{burin2006energy,yao2014many}. 
This so-called critical regime is realized by dipolar spins in 3D, where a combination of power-law interactions, dimensionality, and disorder govern the microscopic dynamics~\cite{anderson1958absence,levitov1990delocalization,kravtsov2012levy}.
Such systems have long been explored in the context of nuclear magnetic resonance spectroscopy, where a wide variety of techniques have been developed to effectively engineer and control spin dynamics~\cite{slichter1990principles,vega1985relaxation,robyr1989radio,wu1989selective,zhang1992polarization}. Despite this, the direct observation of slow, critical dynamics in the presence of strong, controllable disorder remains an outstanding challenge, owing in particular to difficulties in preparing a low-entropy nuclear spin state.

Our approach to the realization and study of critical dynamics makes use of disordered, strongly interacting electronic spin impurities associated with Nitrogen Vacancy (NV) centers in  diamond. 
In particular, we study the thermalization dynamics of an initially polarized spin ensemble  coupled to a bath of unpolarized spins (Fig.~1A).
We directly probe the spin decay dynamics as a function of disorder and  identify a regime of critically slow relaxation.
Our experimental system consists of a dense ensemble of NV centers under ambient conditions (Fig.~1B). 
Each NV center constitutes a $S=1$ electronic spin with three internal states $\ket{m_s = \pm1}$ and $\ket{m_s=0}$, which can be initialized, manipulated and optically read out (Fig.~1C). The NV concentration in our sample is approximately $45$~ppm~\cite{SOM}, yielding an average NV-to-NV separation of $5$~nm and a corresponding typical dipolar interaction strength $J\sim(2\pi)~420$~kHz; crucially, this is significantly faster than the typical spin coherence times~\cite{balasubramanian2009ultralong}. 
Owing to lattice strain and an abundance of other paramagnetic impurities (consisting mainly of P1 centers and $^{13}$C nuclear spins), our system is also characterized by strong disorder; for each NV, this disorder arises from effective random fields generated by its local environment. The magnitude of the disorder $W$ can be directly extracted from an electron spin resonance (ESR) measurement (Fig.~1D), yielding a Gaussian distribution with a standard deviation $W \approx(2\pi)~4.0$~MHz \cite{SOM}. Such an environment can also undergo its own dynamics (e.g. due to spin diffusion among the impurities), resulting in possible variation of the effective random fields over time.

Each NV center in the ensemble can be oriented along any of the four crystallographic axes of the diamond lattice. Different projections of an external magnetic field naturally lead to distinct energy splittings and define four unique NV groups, \{A, B, C, D\}, which can be individually addressed and controlled in a finite B-field via resonant microwave radiation.
By tuning the direction of the magnetic field, one can modify the number of spectrally overlapping groups (e.g. groups B, C in Fig.~1D) and hence the effective density of spins. To directly probe the interaction strength within our system, we perform a double electron-electron resonance (DEER) measurement between two spectrally separated NV groups, A and B (Fig.~2A, bottom inset). In this measurement the spin echo protocol decouples group A from slowly varying magnetic noise. However, the additional $\pi$-pulse on group B after half of the total evolution ensures that the dephasing induced by interactions between the two groups is not decoupled. As depicted in Fig.~2A, by comparing the decay of group A with and without the $\pi$-pulse, this measurement allows us to extract the interaction strength $\sim(2\pi)~420$~kHz~\cite{SOM}. By tuning additional NV groups into spectral resonance, we can confirm that the spin dynamics are dominated by interactions. As a function of the number of resonant groups, $\nu$, we find a total dephasing rate, $\gamma_T = \gamma_{b} + \nu \gamma_{0}$, with $\gamma_{b} \approx 0.9$~MHz and  $\gamma_{0} \approx 0.4$~MHz, consistent with 45~ppm NV center density (Fig.~2A inset)~\cite{SOM}. The linear dependence of $\gamma_T$ on $\nu$ suggests that the dephasing is dominated by coherent interactions, whose strength is proportional to the density of resonant NV groups.

Central to our thermalization experiments is the ability to tune both the disorder strength and interactions. This is achieved by using spin-locking and Hartman-Hahn (HH) resonances, both of which rely upon continuous microwave driving resonant with the $\ket{m_s = 0} \rightarrow \ket{m_s = -1}$ transitions of the respective NV groups~\cite{hartmann1962nuclear,belthangady2013dressed}.
For excitation with Rabi frequency $\Omega$, this defines  a  ``dressed-state'' basis, $\ket{\pm} \approx (\ket{m_s = 0} \pm \ket{m_s = -1}) /\sqrt{2}$~(Fig.~2B).
In the rotating frame, the energies of these two states are split by the effective on-site potential $\sqrt{\Omega^2 + \delta_i^2}$, where $\delta_i$ is the local disorder potential for spin $i$ (of order $W$). In the limit of strong driving $\Omega \gg \delta_i$, we obtain an effective disorder potential $\tilde{\delta}_i$ with the reduced width $W_\textrm{eff} \sim W^2/\Omega$, allowing us to tune the disorder by simply adjusting the Rabi frequency.
For spin-locking, we initialize NVs along the $\hat{x}$-axis via a $\pi/2$-pulse around the $\hat{y}$-axis, thereby polarizing spins in the dressed-state basis. Following coherent driving around the $\hat{x}$-axis for time $\tau$, an additional $\pi/2$-pulse allows the measurement of the polarization in this basis. 
Figure~2C shows a spin-lock experiment performed at two Rabi frequencies.
In comparison to the spin coherence time obtained from a spin-echo measurement, we observe a dramatic enhancement of the lifetime.
We find that the lifetime is limited by interactions with short-lived spins in our system, which is suppressed by increasing $\Omega$~\cite{SinkPaper}.
Thus, spin-locking enables us to prepare a single group of polarized NVs with tunable disorder and long lifetime.

To control interactions, we utilize a HH resonance permitting cross-polarization transfer between the two spin ensembles ~\cite{hartmann1962nuclear,belthangady2013dressed}. To identify the HH resonance condition, two groups of NVs are initialized along $+\hat{x}_A$ (group A) and $-\hat{x}_B$ (group B), and spin-locked along $+\hat{x}_A$ and $+\hat{x}_B$ with Rabi frequencies $\Omega_A$ and $\Omega_B$, respectively (Fig.~3A).
This prepares two oppositely polarized spin ensembles in the dressed-state basis with energy splittings $\Omega_A$ and $\Omega_B$.
The interaction between the groups results in spin exchange and leads to a resonant cross-relaxation when $\Omega_A = \Omega_B$ (HH condition). 
Figure~3B depicts the results of a spin-lock measurement on group A as a function of $\Omega_B$, revealing a sharp resonance with a linewidth significantly narrower than the on-site disorder strength $W$.
The linewidth of this resonance can be monitored as a function of the common Rabi frequency $\Omega = \Omega_A = \Omega_B$, showing a strong decrease for higher $\Omega$ caused by a reduction of the effective disorder $W_\textrm{eff}$ (Fig.~3B inset). 

This method allows us to probe the controlled thermalization dynamics with tunable disorder. 
To this end, we investigate the dynamics of an initially polarized spin sub-ensemble (group A) in HH resonance with another, unpolarized sub-ensemble (group B).
Physically, this situation corresponds to the thermalization of a polarized spin ensemble in contact with a spin bath held at infinite effective temperature.
To extract the coherent thermalization dynamics, we normalize the polarization decay with a sufficiently detuned HH measurement~\cite{SOM}, wherein we observe a decay profile that fits neither a diffusive power law ($\sim t^{3/2}$) nor a simple exponential (Fig.~3C).
By varying the driving strength $\Omega$, we find that the polarization decays faster for larger $\Omega$, consistent with a smaller effective disorder (Fig.~3D).
Interestingly, the functional profile of the decays is consistent with power laws for over a decade, followed by accelerated relaxation at late times.


To understand these observations, we turn to a theoretical description of our system.
Spin dynamics are governed by the interplay between disorder and long-range dipolar interactions.
Working in the dressed state basis with the quantization axis along $\hat{x}$, we find that the form of this interaction depends on whether spins reside in the same or in distinct groups. 
For spins in different groups (A and B), dipolar interactions naturally lead to spin exchange, $H_{AB} = \sum_{i\in A, j\in B} J_{ij}/r_{ij}^3 ~(\sigma_i^+ \sigma_j^- +\sigma_i^- \sigma_j^+)$, where $r_{ij}$ is the distance between spins, $J_{ij}$ is the orientation dependent coefficient of the dipolar interaction with typical strength $J_0 = (2\pi)~52$~MHz$\cdot$nm$^3$, and $\vec{\sigma}$ are spin-1/2 operators with $\sigma_i^{\pm} = \sigma_i^{y} \pm i \sigma_i^{z} $ ~\cite{SOM}.
However, for spins in the same group, the $S=1$ nature of the NV centers and energy conservation in the rotating frame lead to an absence of spin exchange ~\cite{SOM}; rather, the coupling between spins takes the form of an Ising interaction, $H_{A(B)} = \sum_{i,j\in A(B)} J_{ij}/r_{ij}^3~\sigma_i^{x} \sigma_j^{x} $. 
Thus, when initially polarized, a spin may depolarize only through exchange with spins of the other group.
Specifically, in the limit of strong disorder, one expects the dynamics to be dominated by rare resonant exchange processes between the two groups.
To describe such dynamics, we consider a simplified model, where a single group A excitation is located at the center of an ensemble of group B spins (Fig~4A). 
The dynamics of this excitation are captured by an effective Hamiltonian, 
\begin{equation}
H_{\textrm{eff}} = \sum_i \tilde{\delta}_i \sigma_i^x - \sum_{ij} \frac{ J_{ij}}{r_{ij}^3} (\sigma_i^+ \sigma_j^- + h.c.).
\end{equation}
where $\tilde{\delta}_i = \sqrt{\Omega^2 + \delta_i^2} -\Omega$ is the effective quenched disorder potential. 
While this single-particle model neglects the many-body nature of our experiments such as intra-group Ising interactions and complex dynamics of group B excitations, it captures the key features of slow relaxation in critical systems; however these additional features will be necessary to accurately describe the long time thermalization behavior. 

To characterize the spin decay dynamics governed by $H_{\textrm{eff}}$, we calculate the survival probability, $P(t)$, of the excitation via a simple resonance counting analysis. For a given disorder realization, this resonance counting proceeds as follows. Two spins  at sites $i$ and $j$ are on resonance at time  $t$ if: (1) their energy mismatch is smaller than their dipolar interaction strength, $|\tilde{\delta}_i - \tilde{\delta}_j | < \beta J_0 / r_{ij}^3$ ($\beta$ is a dimensionless constant of order unity), and (2) the interaction occurs within the time-scale $t$, $J_{ij} / r_{ij}^3 > 1/t$. 
$P(t)$ is approximately given by the probability of having found no resonances up to time $t$ or equivalently up to distance $R(t)$\,$\equiv$\,$(J_0 t)^{1/3}$\,\cite{SOM}.
This probability can be computed as the product of probabilities of having no resonant spins at any\,$r$,
\begin{equation}
\label{eqn:single_particle_power_law}
P(t) =  \prod_{r}^{R(t)} \left( 1 - 4 \pi n r^2 dr \frac{\beta J_0 / r^3}{W_\textrm{eff}}\right) \propto t^\frac{-4\pi n \beta J_0}{3W_\textrm{eff}}.
\end{equation}
$P(t)$ exhibits power-law decay with a disorder dependent exponent $\eta = 4\pi n \beta J_0/(3W_\textrm{eff})$, where $n$ is the density of spins that are oppositely polarized to the central excitation. 
This sub-exponential relaxation is the essence of the slow critical dynamics predicted by Anderson \cite{anderson1958absence}. 
Such single-particle power-law relaxation is also consistent with results obtained from random-banded matrix theory~\cite{mirlin1996transition,kravtsov2012levy} 
and is numerically verified for up to $N=10^4$ spins \cite{SOM}.

A detailed comparison of our experimental observations with these theoretical predictions is summarized in Fig.~4.
In order to quantify the slow dynamics, we take subsets of our depolarization time trace over half-decade windows, fit the data to power laws, and  extract the resulting exponents. Varying the starting time of the windows, we find that the extracted exponents remain constant up to a long time $T^* \gg 1/J$, beyond which they increase, indicating the deviation of the thermalization dynamics from a simple single particle prediction (Fig.~4B). 
Interestingly, the exponents scale linearly with the inverse effective disorder, as predicted by the counting argument (Fig.~4C).
Furthermore, we find that their values are in excellent agreement with our theory based on numerical simulations of a single-particle Hamiltonian~\cite{SOM}.

At late times ($t > T^*$), the observed decay accelerates and deviates significantly from the power law. This is natural since the effects of multi-particle interactions cannot be neglected when a significant fraction of spins have already undergone depolarization.
In particular, intra-group Ising interactions among randomly positioned spins $\delta^{I}_i$\,$\equiv$\,$\sum_j\,J_{ij}/r_{ij}^3$\,$\langle \sigma^x_j \rangle$ may behave as an additional disorder that change in time with characteristic strength $J/4$\,$\sim$\,$(2\pi)$\,$105$\,kHz.
Additionally, weak coupling to the bath may also give rise to corrections to our single particle model.

To understand this behavior, we modify our theoretical model by considering the time dependence of the on-site disorder potential. More specifically, we assume that the disorder potential consists of both static and dynamic parts with standard deviations $W_s$ and $W_d$. The dynamic disorder is assumed to vary at a slow rate $1/\tau_d$.
The survival probability $\tilde{P}(t)$ can then be computed using a modified resonance condition where two spins are considered resonant at time $t$, if at any prior time $t'$, their energy mismatch is smaller than their dipolar interaction strength,  $|\tilde{\delta}_i (t') - \tilde{\delta}_j(t')| < \beta J_{0}/r^3_{ij}$ \cite{SOM}.
Repeating our previous analysis, we find an analytical expression of $\tilde{P}(t)$ which, in the limit of strong quenched disorder ($1/\tau_d < W_d \ll W_s)$, can be approximated by $\tilde{P}(t) \propto e^{-t/T^*} t^{-\eta}$ with $T^* \equiv 3W_s \tau_d/ (4\pi n\beta J_0)$, predicting the deviation from the power law at time scale $T^*$. Intuitively, $1/T^*$ characterizes the rate at which a pair of initially off-resonant spins comes into resonance as the local potentials vary in time. 
Figure 3D shows that $\tilde{P}(t)$ provides an excellent fit to our observation over all time scales.
Both extracted parameters $W_d$\,$\sim$\,$(2\pi)$\,$0.5$\,MHz and $\tau_d \sim40~\mu$s are comparable to the strength of Ising interactions and independently measured NV depolarization time, respectively~ \cite{SinkPaper,SOM}.
This suggests that the dynamical disorder is dominated by intrinsic contributions from Ising interactions, which is related to the predicted thermalization enhancement due to multi-particle resonances and higher order processes~\cite{yao2014many,burin2006energy}.
Moreover, the extracted power-law duration agrees well with the predicted linear dependence of $T^*$ on effective disorder strengths (Fig. 4D).
The quantitative agreement of the functional form $\tilde{P}(t)$, the disorder dependence of power-law exponents and durations, and the extracted values of $W_d$ and $\tau_d$ corroborates our theoretical model describing the microscopic mechanism of thermalization dynamics in a critical system.


We have demonstrated that dense ensembles of NV centers constitute a powerful platform for exploring quantum 
dynamics of strongly correlated many-body systems. Complementary to recent studies of localization in cold atomic systems~\cite{schreiber2015observation,smith2016many}, these spin systems exhibit slow, disorder-dependent relaxation associated with critical thermalization dynamics. The quantitative agreement between the observed spin relaxation and resonance counting demonstrates that the dynamics are dominated by rare resonances.
Moreover, the observed deviations from simple theory reveal the subtle role  that  many-body effects and coupling to the environment can play in such systems. 
These studies can be extended along several directions.
A higher degree of spatial quantum control can be obtained via spin-based sub-wavelength imaging techniques~\cite{maurer2010far}.
Advanced dynamical decoupling can enable the engineering of a broader class of interaction Hamiltonians and the direct measurement of quantum entanglement dynamics~\cite{hauke2016measuring,choi2017hamEng}. 
The use of strong magnetic field gradients or  the fabrication of diamond nanostructures can allow for the exploration of spin dynamics in lower dimensional systems~\cite{burek2012free}, where the existence of many-body localization is still in debate~\cite{burin2006energy,levitov1990delocalization}.
In combination, these directions may enable the study of dynamical phase transitions from localization to thermalization~\cite{langen2015experimental,kaminishi2015entanglement,schreiber2015observation} as well as exotic non-equilibrium phases of matter~\cite{Zhang:2016uw,choi2016observation,potter2016classification}, and open up new opportunities for controlling  such complex interacting systems~\cite{choi2015quantum,yao2015many}.  

\bibliography{biblatex}

\begin{thebibliography}{10}
\bibitem{SOM}
Materials and methods are available as supporting material on science online.
\end{thebibliography}

\noindent \textbf{Acknowledgments}: We thank A. Gali, D. Budker, B. J. Shields, A. Sipahigil, M. Knap, S. Gopalakrishnan and J. Chalker for insightful discussions and N.P. De Leon for fabricating the diamond nanobeam. This work was supported in part by CUA, NSSEFF, ARO MURI, Moore Foundation, Miller Institute for Basic Research in Science, Kwanjeong Educational Foundation, Samsung Fellowship, NSF PHY-1506284, NSF DMR-1308435, Japan Society for the Promotion of Science KAKENHI (No. 26246001), EU (FP7, Horizons 2020, ERC), DFG, SNF, Volkswagenstiftung and BMBF.

\clearpage

\begin{figure}[ht]
\begin{center}
\includegraphics[width=0.8\textwidth]{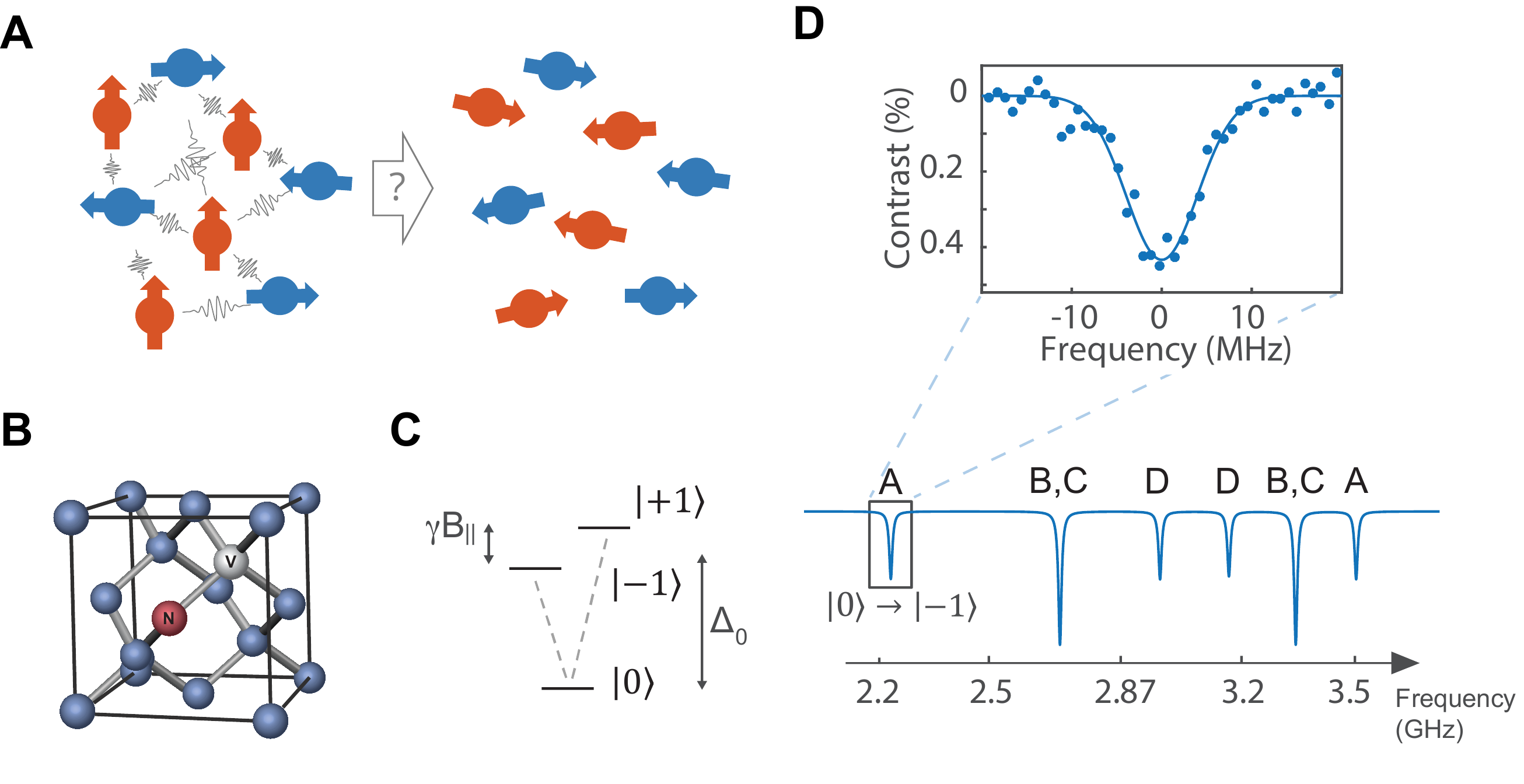}
\end{center}
\caption{\textbf{Experimental System.}
\textbf{(A)} Schematic depicting two groups of spin ensembles interacting via long-range dipolar interactions. An initially polarized system (red arrows) strongly coupled to a bath of unpolarized spins (blue arrows) will eventually thermalize to an unpolarized spin state. \textbf{(B)} The crystallographic structure of diamond contains four possible NV quantization axes, defined by the position of the nitrogen atom and the adjacent vacant lattice site. \textbf{(C)} Simplified NV level scheme showing the spin degrees of freedom in the optical ground state. A large zero-field splitting $\Delta_0 = (2\pi)~2.87$~GHz  in combination with a magnetic field induced Zeeman shift $\gamma B_{\|}$ leads to individual addressability of the spin sub-levels. \textbf{(D)} The lower image shows a simulated ESR scan, revealing the spin transitions of all four NV groups \{A, B, C, D\}. For example, the orientation of the external magnetic field can be chosen in such a way that NV groups B and C experience the same magnetic field projection, leading to spectral degeneracy. The upper figure shows an ESR scan of a single transition of NV spins (blue points). Blue solid line represents a Gaussian fit with standard deviation $W$, corresponding to the average disorder in the sample.}
\end{figure}

\newpage

\begin{figure}[ht]
\begin{center}
\includegraphics[width=0.8\textwidth]{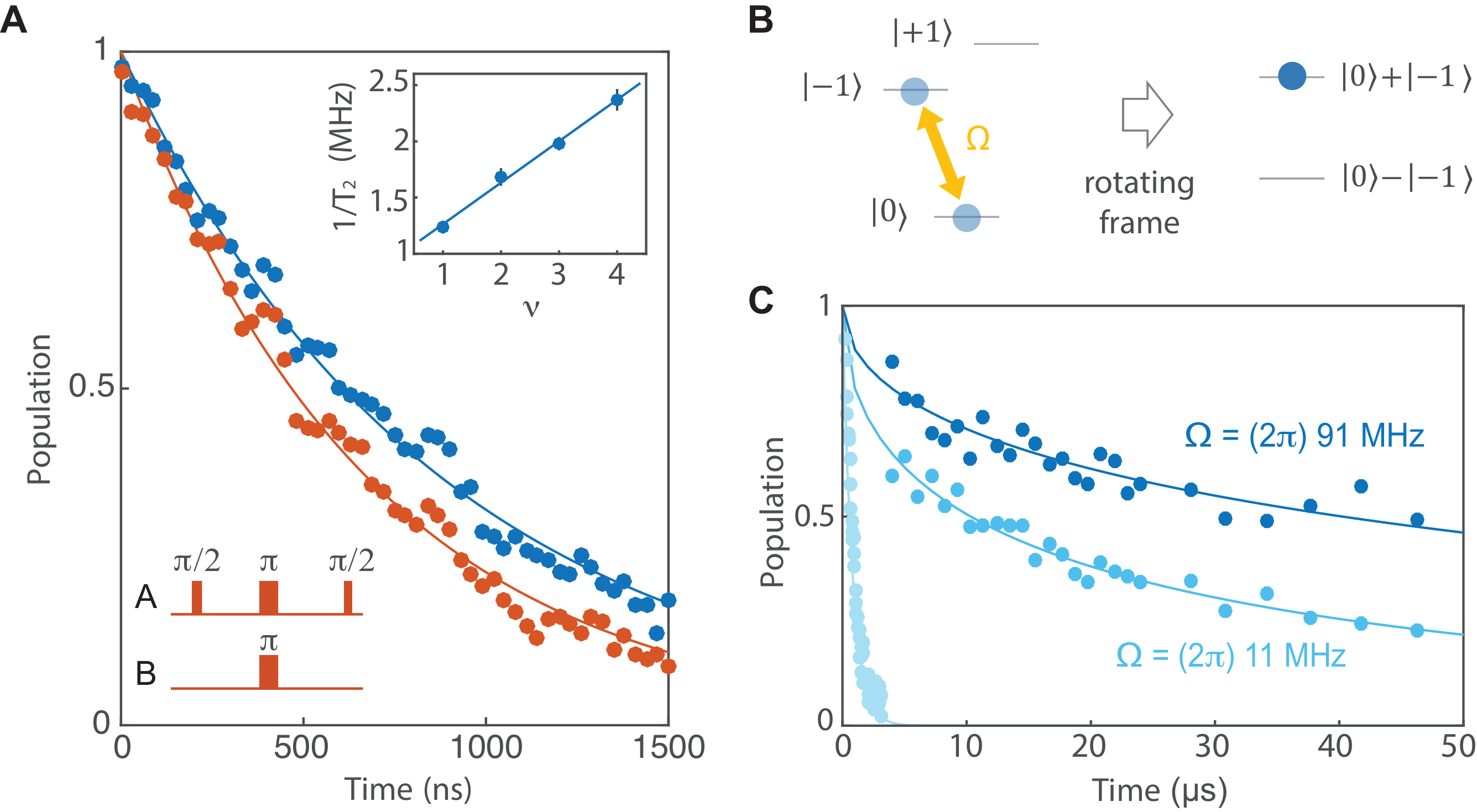}
\end{center}
\caption{\textbf{Interacting Spin Ensemble.} \textbf{(A)} Spin echo on NV group A (blue points) and DEER scan on groups A and B (red points). The bottom left inset illustrates the DEER pulse sequence. Solid lines indicate exponential fits to the data. The inset shows the spin echo coherence time as a function of resonant NV groups (blue points). The blue solid line represents a linear fit to the data. \textbf{(B)} Schematic diagram depicting the NV level scheme during a spin-lock sequence with a driving strength $\Omega$. In the rotating frame, the spin-locking creates a dressed spin basis with the two eigenstates $(|m_s=0\rangle+|m_s=-1\rangle)/\sqrt{2}$ and $(|m_s=0\rangle-|m_s=-1\rangle)/\sqrt{2}$. \textbf{(C)} Spin-lock coherence decay for low (light blue points) and high (dark blue points) CW driving power, showing significant extension beyond the spin echo decoherence (gray-blue points). The decay curves are fitted to a stretched exponential function $\exp{[-\sqrt{t/T}]}$~\cite{SinkPaper}.}
\end{figure}

\newpage

\begin{figure}[ht]
\begin{center}
\includegraphics[width=0.9\textwidth]{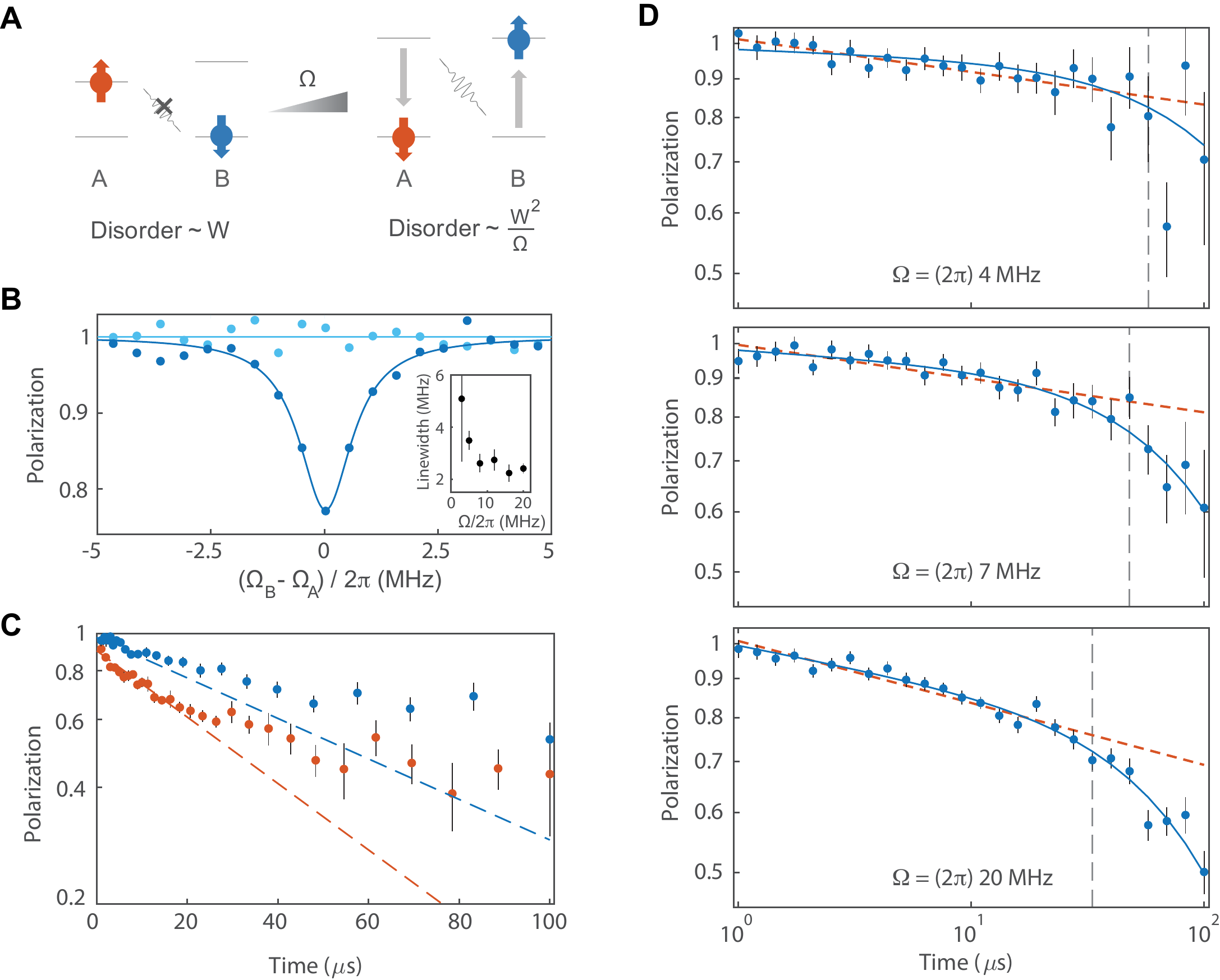}
\end{center}
\caption{\textbf{Spin Cross-Relaxation.} \textbf{(A)} Schematic diagram depicting two oppositely polarized groups of spins (A and B) in the dressed state spin basis. Microwave fields with a common driving strength $\Omega$ are independently applied to the respective group parallel to their spin polarization directions. Under spin-locking, the effective disorder reduces from the natural disorder $W$ with increasing $\Omega$, thereby enhancing the rate of resonant spin exchange. \textbf{(B)} Population of group A as a function of driving power of group B, showing the HH resonance (dark blue points). Light blue data shows spin-lock coherence without driving of other groups. The corresponding solid curves represent a Lorentzian and constant fit to the data. Inset shows the HH resonance linewidth as a function of applied Rabi frequency. \textbf{(C)} Polarization dynamics of group A interacting with an oppositely polarized (red) and unpolarized (blue) spin bath, group B, at the HH condition with $\Omega$ = (2$\pi$) 9 MHz as a function of evolution time (on a semi-log scale). The polarized spin bath leads to faster polarization decay (see Supplementary Materials). Dashed lines represent an exponential decay, illustrating significant deviation at long times. \textbf{(D)} Double-logarithmic plot of polarization decay of group A under HH conditions with unpolarized group B for different driving strengths. Dashed red lines are power-law fits to the data in the time window up to the vertical line. Curved solid lines are the fits to our theory model including time-dependent disorder. All errorbars correspond to 1~$\sigma$.}
\end{figure}

\newpage

\begin{figure}[ht]
\begin{center}
\includegraphics[width=0.7\textwidth]{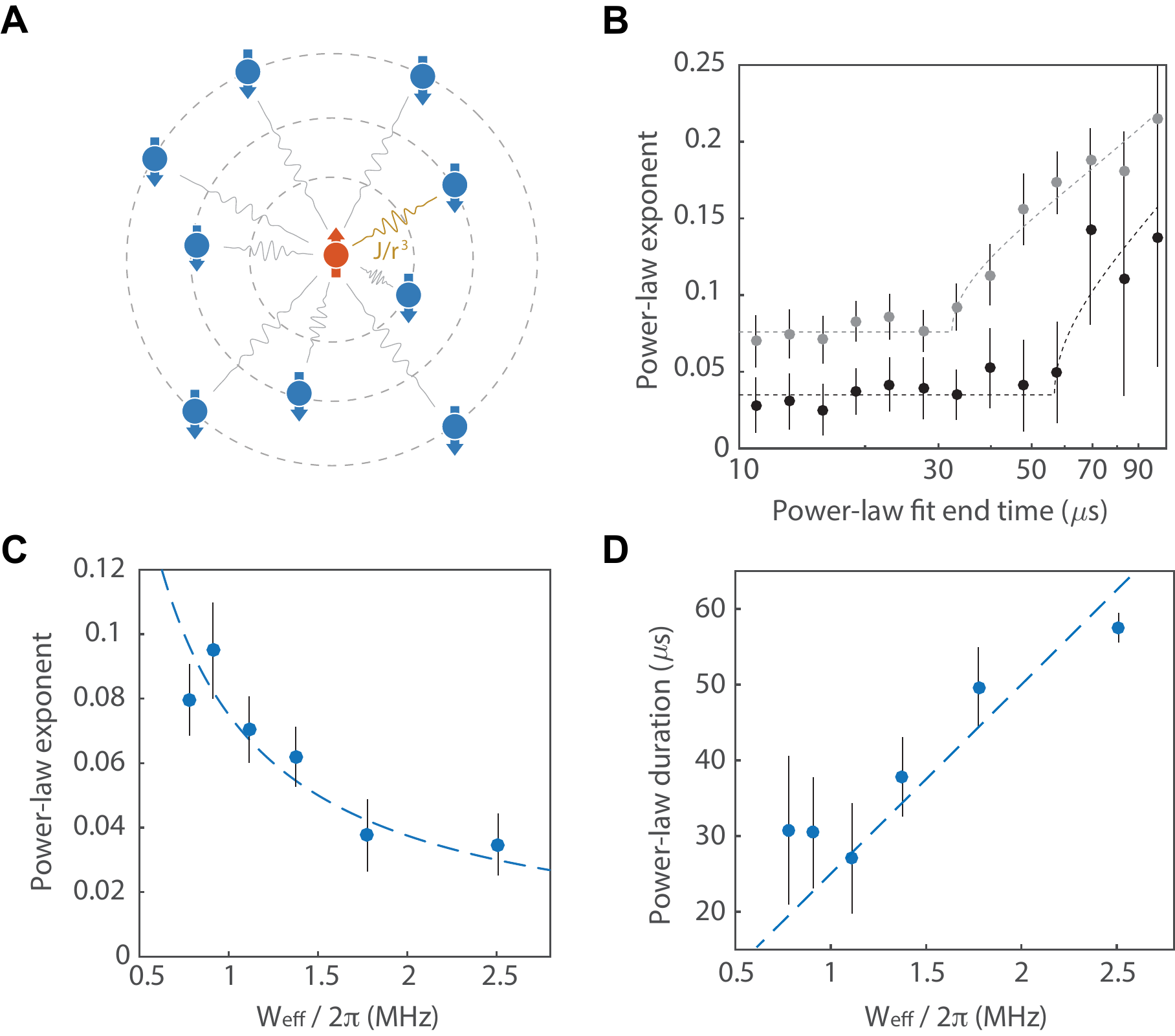}
\end{center}
\caption{\textbf{Understanding Thermalization Dynamics.} \textbf{(A)} Schematic diagram of single particle resonance counting argument predicting a power-law decay profile. \textbf{(B)} Variation of power-law exponents extracted from a subset of data, consisting of seven subsequent points, swept from the beginning to the end of thermalization time traces. Black and gray data corresponds to the case of $\Omega = (2\pi)$ 4 and 9 MHz, respectively. Dotted lines corresponds to phenomenological fits, identifying the duration over which the power-law exponents remain constant. \textbf{(C)} Exponent of the power-law decay of group A polarization as a function of effective disorder $W_\textrm{eff}$. Dashed line corresponds to a theoretical prediction based on the single particle resonance counting. \textbf{(D)} Duration of power-law dynamics extracted for various strengths of effective disorder $W_\textrm{eff}$. Dashed line corresponds to a theoretical prediction based on a refined resonance counting including time-dependent disorder. All errorbars correspond to 1~$\sigma$.}
\end{figure}
\end{document}


\baselineskip24pt

\setcounter{tocdepth}{4}
\setcounter{secnumdepth}{4}

\maketitle

\tableofcontents

\section{Materials and Methods}

\subsection{Sample Fabrication}

The diamond sample used in this work (type-Ib, $\sim$4~mm in diameter) was grown via high pressure and high temperature (HPHT), at 5.5~GPa and 1350~$^o$C, using a Fe-Co alloy as a solvent. The main source of paramagnetic impurities was provided by substitutional nitrogen atoms in the neutral charge state (P1 centers) at a concentration of $\sim$100 ppm. A diamond plate of thickness $\sim$1~mm was obtained via laser cutting and polishing.
To obtain NV centers, high energy electron irradiation was performed at $\sim$2~MeV with a flux of 1.3-1.4$\cdot$10$^{13}$~e$\cdot$cm$^{-2}\cdot$s$^{-1}$ and in-situ annealing at 700-800~$^o$C up to a total fluence of 1.4$\cdot$10$^{19}$~cm$^{-2}$ (total time of 285~hrs). Additional annealing at 1000~$^o$C for 2~hrs in vacuum was performed after half as well as after the full irradiation time.
This process resulted in the diamond with NV centers of a concentration $\sim$45~ppm, corresponding to $\sim$5~nm of average separation and $ \sim (2\pi)$~420~kHz dipole-dipole interaction strength. To control the region of optical excitation, we used angle etching to create a beam-shaped piece of diamond, of 20~$\mu$m length and $\sim$300~nm width, and transferred it onto our coplanar waveguide~\cite{burek2012free}.

\subsection{Optical Setup}

\begin{figure}[h!]
\begin{center}
\includegraphics[width=1\textwidth]{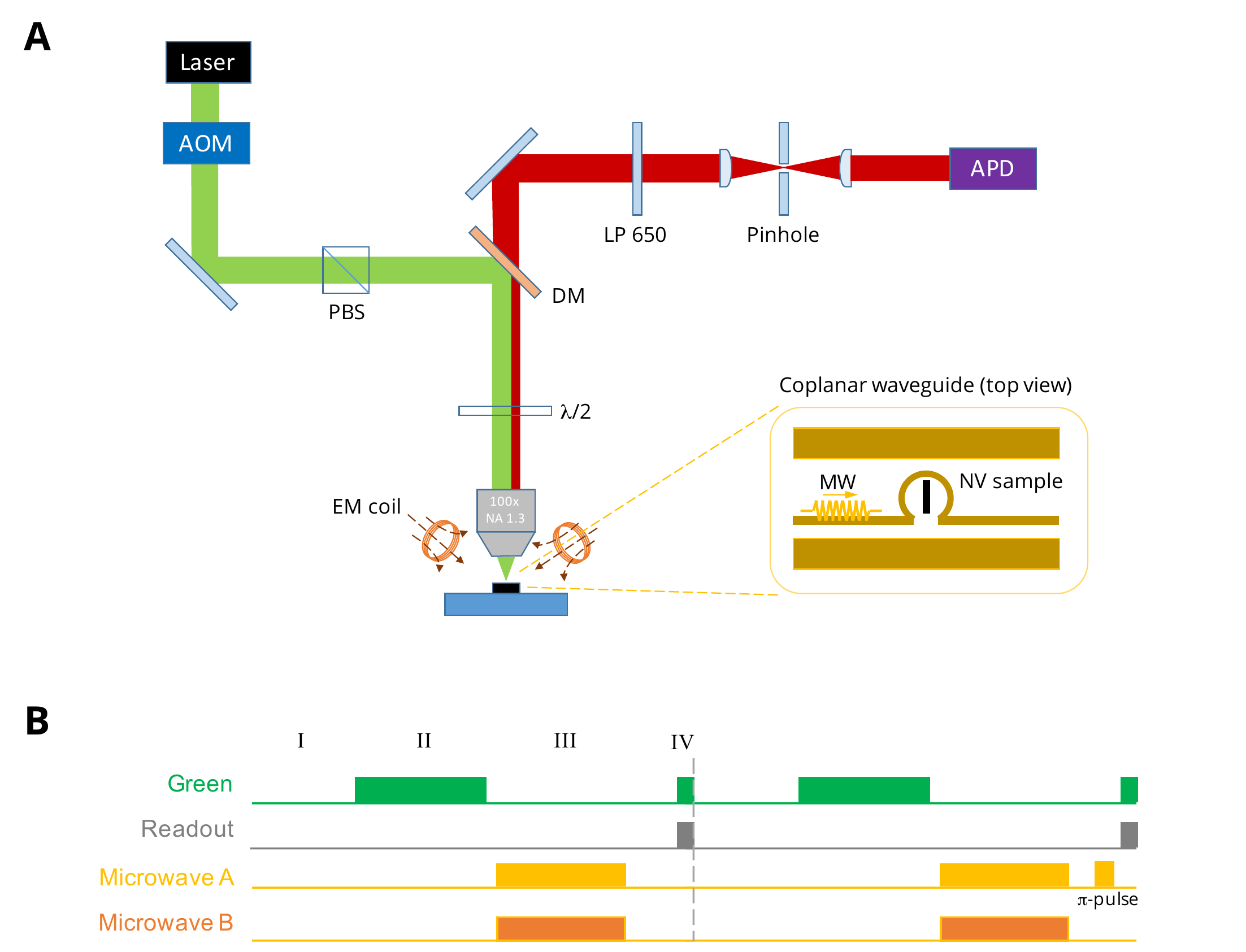}
\end{center}
\caption{\textbf{Schematic of the Optical Setup and the Pulse Sequence.} (\textbf{A}) Green and red lines indicate the optical paths (excitation: green, collection: red). An acousto-optic modulator (AOM) is used to control green laser duration. A dichroic mirror (DM) spectrally filters out the fluorescence from NV centers for electronic spin state readout. A 650~nm long pass filter additionally helps to filter fluorescence emission, corresponding to the phonon sideband (PSB) of NV centers. A 5-$\mu$m pinhole is used in combination with a single photon counting avalanche photodiode (APD) to achieve confocal detection. A polarizing beam splitter (PBS) is used to polarize the excitation beam. With the addition of a $\lambda/2$ waveplate we get control over the incident green polarization onto the diamond sample. The sample is placed ontop of a coplanar microwave (MW) structure in the shape of an omega (inset). Three electromagnetic coils are used to create a static magnetic field up to $\sim$300~Gauss in an arbitrary direction. (\textbf{B}) Typical experimental sequence used to measure NV dynamics. I: charge equilibration; II: spin polarization; III: experimental sequence; IV: spin readout.}
\label{fig:OpticalSetup}
\end{figure}

As shown in Fig.~\ref{fig:OpticalSetup}A, the optical setup consists of a home-built confocal microscope with a Nikon Plan Fluor 100x oil immersion objective (NA = 1.3). The sample is mounted on a xyz-piezoelectric stage in the focal plane of the microscope. Excitation of the ensemble of NV centers is performed by illuminating a green laser ($\lambda =$ 532~nm) with average power less than 50~$\mu$W. Short laser pulses are generated by an acousto-optic modulator (AOM) from Isomet in a double pass configuration. The $\lambda/2$-waveplate at the objective allows the control over the polarization of excitation light. NV centers emits fluorescence into the phonon sideband (630-800~nm), which is isolated from the excitation laser by a dichroic mirror. An additional 650~nm long-pass filter further suppresses the detection of unwanted signal. After passing a pinhole the collection beam is then focused onto a single photon counting avalanche photodiode (APD) to achieve detections with confocal resolution. 

To probe the spin dynamics over time, we used a pulse sequence illustrated in Fig. S1B. We repeat the same pulse sequence twice, but include an extra $\pi$-pulse right before the read-out at end of the second sequence. The photon-count difference between the two read-outs allows us to measure the NV polarization, while being insensitive to changes in the background fluorescence due to charge dynamics~\cite{SinkPaper}.

\subsection{Microwave Setup}

\begin{figure}[h!]
\begin{center}
\includegraphics[width=1\textwidth]{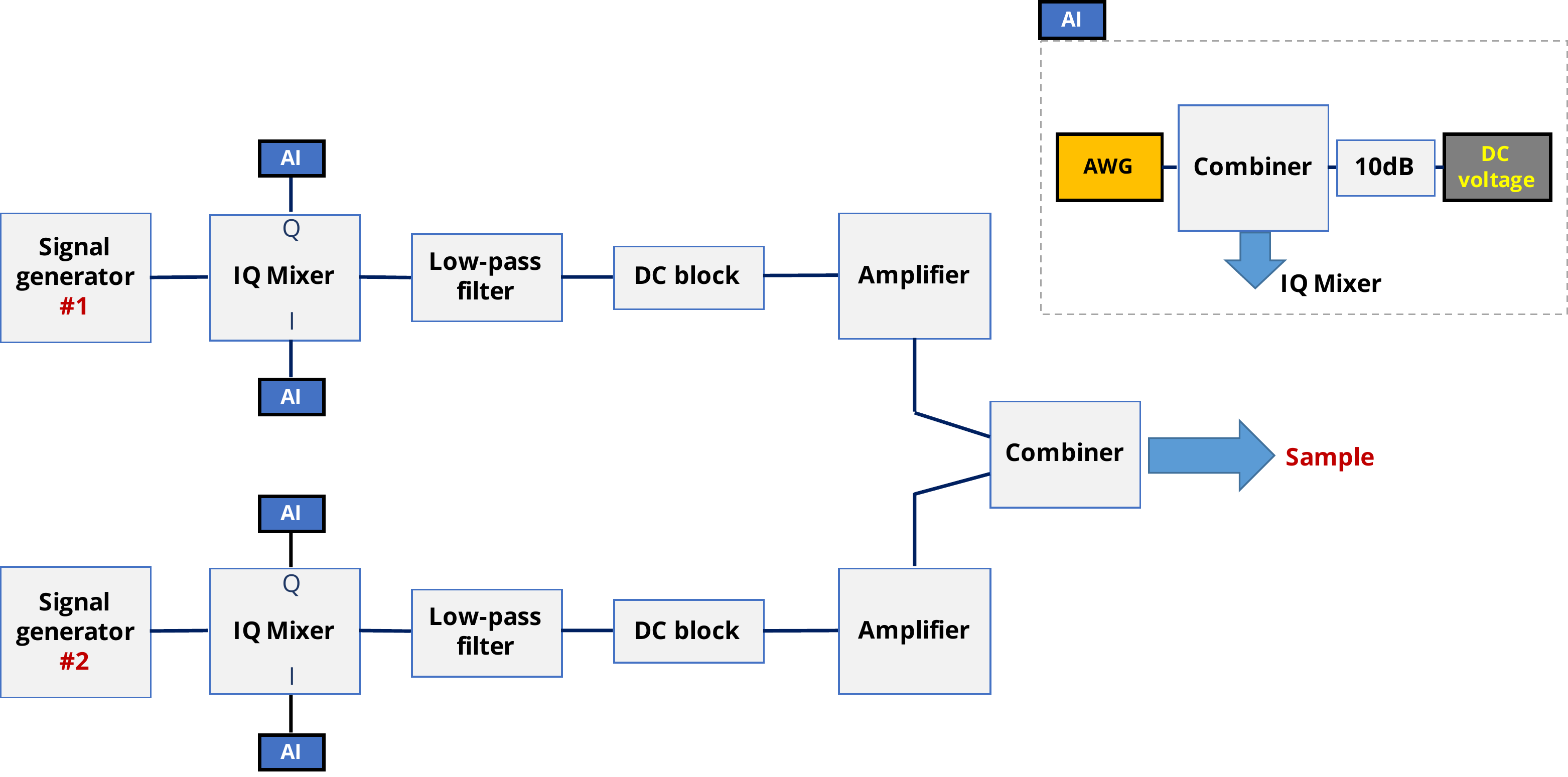}
\end{center}
\caption{\textbf{Schematic of the Microwave Control Setup.} Two sets of independent microwave circuits are used to achieve full control over two separate groups of NV centers at different transition frequencies. A 3~GHz low-pass filter suppresses unwanted higher-order harmonics. The two microwave paths are separately amplified to avoid saturation and then combined and sent to the diamond sample. In order to  precisely control the microwave pulse length as well as phase, each path is sent through an IQ mixer controlled by an arbitrary waveform generator (AWG) output. The inset shows the detailed configuration of analog inputs connected to the IQ mixers used to define microwave pulse length and phase. In order to finely tune the voltage offset of the I and Q port, to achieve high isolation, a DC voltage source is combined with the AWG signal. The addition of an attenuator allows the suppression of unwanted reflections. }
\label{fig:MWSetup}
\end{figure}

To coherently control the electronic spin states of NV centers we deliver microwaves to the sample through an impedance-matched coplanar waveguide fabricated on a glass coverslip.
An omega-shaped microstructure (with a inner diameter 20~$\mu$m) at the center of the waveguide allows us to achieve Rabi frequencies up to $\sim$(2$\pi$) 100~MHz. In Fig.~S2, we illustrate the schematic diagram of the microwave control system. 
In order to have full control over two groups of NV centers with different transition frequencies, we employ two independent microwave circuits.
In each circuit, a RF signal generator (Rohde \& Schwarz SMIQ06B) produces 
the main driving frequency;
an IQ mixer (Marki IQ1545LMP) generates pulsed signals;
a low-pass microwave filter (Mini-Circuits VLF-3000+) suppresses unwanted higher-order harmonics of fundamental frequencies;
and a DC block (Picosecond 5501a) additionally isolates the signals from low-frequency noises.
After separately amplified (ZHL-16W-43+), two RF signals are then combined by a power combiner (Mini-Circuits ZFRSC-42-S+) and delivered to our sample.
The inset of Fig.~\ref{fig:MWSetup} depicts the detailed configuration of analog inputs (AI) connected to the IQ mixers.
An arbitrary waveform generator (The Tektronix AWG7052) defines the duration and the phase of the pulses with a temporal resolution of 1 ns.
For fine tuning of the voltage offset on the I and Q ports, a DC voltage is applied to the AWG signal.
The addition of a 10-dB attenuator between the voltage source and the combiner suppresses unwanted reflections (see inset of Fig. S2).

\subsection{Magnetic Field Setup}

\begin{figure}[h!]
\begin{center}
\includegraphics[width=1\textwidth]{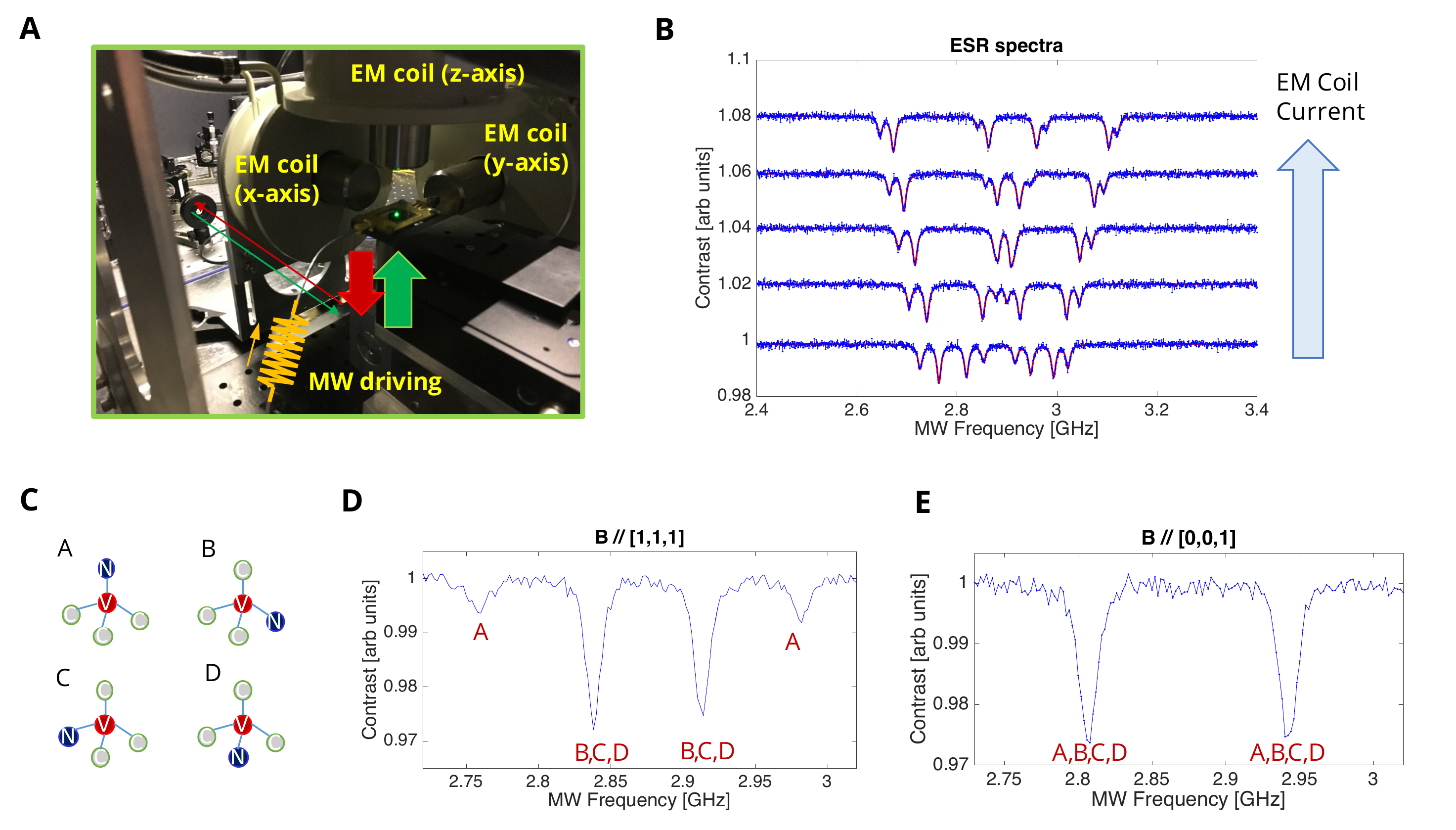}
\end{center}
\caption{\textbf{Magnetic Field Calibration and Control.} \textbf{(A)} Three electromagnetic (EM) coils are located in the vicinity of the diamond sample in order to provide an external magnetic field (B-field) in an arbitrary direction with an amplitude up to $\sim$300~Gauss. \textbf{(B)} To calibrate the coil's magnetic field, electron spin resonance (ESR) spectra are recorded for different values of coil currents. \textbf{(C)} The diamond lattice allows for four different crystallographic orientations of NV centers. The different groups A, B, C, and D of NV centers are characterized by their N-V axis orientations, i.e., A = [111], B = [$\bar{1}\bar{1}$1], C = [1$\bar{1}\bar{1}$], and D = [$\bar{1}$1$\bar{1}$]. \textbf{(D)} Measured ESR spectrum for the B-field aligned along the [111] direction. Group A exhibits the largest Zeeman splitting (highest projected $B_{||}$) because the spin quantization axis of group A is parallel to the chosen B-field. \textbf{(E)} Measured ESR spectrum for the B-field aligned along the [001] direction. Due to the [100] cutting direction of the diamond, all 4 NV groups form the same angle to the surface. With the external B-field being perpendicular to the sample surface, this leads to groups A-D having degenerate B field projections. }
\label{fig:BfieldSetup}
\end{figure}

For an external magnetic field, we use three water-cooled electromagnetic (EM) coils,
which can provide a B-field up to $\sim$300~Gauss in an arbitrary orientation (see Fig.~\ref{fig:OpticalSetup}A and \ref{fig:BfieldSetup}A). 

As shown in Fig. S3B, we calibrate the magnetic field by recording electron spin resonance (ESR) spectra at various values of currents in the coils; since the Hamiltonian of a NV center in the presence of a magnetic field is known, the magnetic field  at the position of our sample can be extracted from transitions frequencies of NV centers. For this process we utilize all four groups $\{A, B, C, D\}$ of NV centers oriented in different crystallographic axes of diamond lattice, e.g., A = [111], B = [$\bar{1}\bar{1}$1], C = [1$\bar{1}\bar{1}$], and D = [$\bar{1}$1$\bar{1}$] (see Fig. S3C).
Fig. \ref{fig:BfieldSetup}D shows an ESR spectrum when the B-field is aligned along [111] direction; group A exhibits the largest Zeeman splitting, while the other groups B, C, and D become degenerate.
In Fig.~\ref{fig:BfieldSetup}E, the direction of an external B-field is perpendicular to the sample surface, i.e., B $\|$ [001], resulting in four degenerate groups.

\section{Characterization of Experimental System}

\subsection{On-site Potential Disorder}

The ESR linewidth of an NV ensembles is influenced by multiple factors. To discuss and estimate their contributions we introduce the ground state hamilitonian of the electronic spin state of a single NV center: 
\begin{align}
H = \left( \hbar\Delta_0 + d_{||} E_{||}^z \right) S_z^2 + \gamma_{NV} (\vec{S} \cdot \vec{B}) - d_{\bot} \left[ E_{\bot}^x \left( S_x S_y + S_y S_x \right) + E_{\bot}^y \left( S_x^2 - S_y^2 \right) \right], 
\end{align} 
where $S_x$, $S_y$ and $S_z$ denote the spin-1 matrices and $\hbar$ the reduced Planck constant; $\Delta_0 \approx (2\pi) 2.87$ GHz, $\gamma_{NV}$ = $(2\pi)~$2.8 MHz G$^{-1}$, $d_{\parallel}$ = $(2\pi)~$0.35 Hz cm V$^{-1}$ and $d_{\bot}$ = $(2\pi)~$17 Hz cm V$^{-1}$ are the zero field splitting, the gyromagnetic ratio, axial and perpendicular components of the ground triplet state permanent electric dipole moment of a NV center \cite{van1990electric}. 
$B_{\parallel(\bot)}$ and $E_{\parallel(\bot)}$ are projection of the effective magnetic and electric field parallel (perpendicular) to the NV axis. To a leading order, we ignore the effect of the perpendicular magnetic field noise $\delta B_{\bot}$, since it influence less on the spin coherence than the parallel one $\delta B_{\parallel}$, owing to the large zero field splitting. 

\begin{figure}[h!]
\begin{center}
\includegraphics[width=0.75\textwidth]{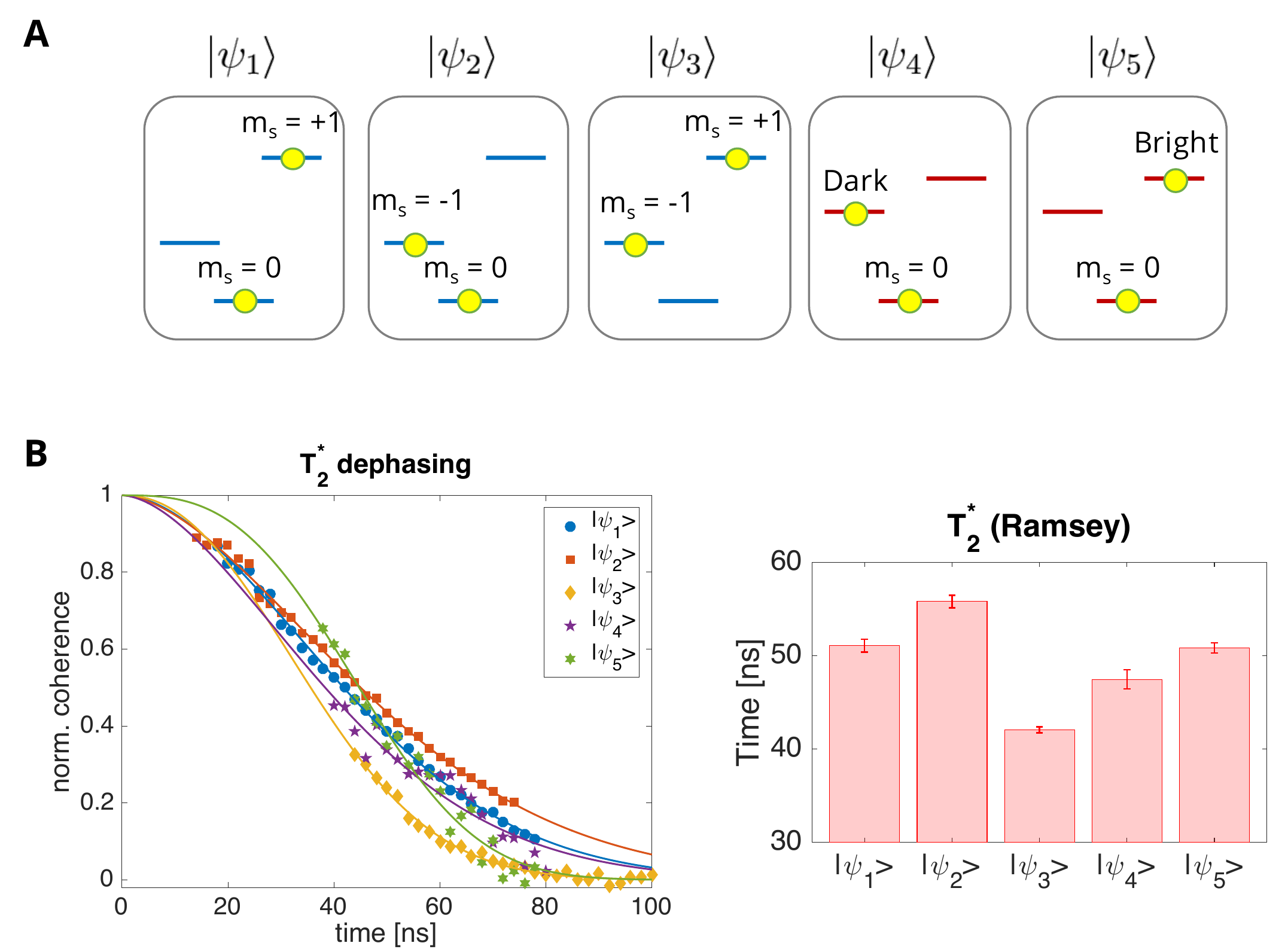}
\end{center}
\caption{\textbf{$T_2^*$ and $T_2$ Measurements of different Basis States.} \textbf{(A)} Different initial conditions used for coherence measurements. By aligning the magnetic field parallel (perpendicular) to the NV axis, the eigenbasis for the spin state of NV centers becomes \{$|m_s = 0\rangle, |m_s = +1\rangle, |m_s = -1\rangle$\} (\{$|m_s = 0\rangle$, $|Dark\rangle$, $|Bright\rangle$\}), where Bright and Dark states are defined as even and odd combination of the original bare spin states $|m_s = -1\rangle$ and $|m_s = +1\rangle$. \textbf{(B)} Ramsey spectroscopy data and extracted decay timescale for different initial states.}
\label{fig:Dephasing}
\end{figure}

To account for effects of the local NV environment we include in $B_{||}$ and $E_{||(\bot)}$ on-site potential disorders originating from randomly distributed magnetic fields due to nuclear spins (i.e. $^{13}$C or $^{14}$N) and paramagnetic impurities (i.e. P1 centers) as well as fields caused by local electric fields and lattice strain. To quantify the different contributions to the ESR linewidth, we conduct Ramsey spectroscopy in distinct basis states as listed in Fig. \ref{fig:Dephasing}A. Since each basis has a well defined sensitivity to different physical noise sources, our Ramsey measurements provide insight into the local environment of the NV centers. Table \ref{tab:disorder} lists the effects of magnetic and electric field noise on free induction decay of several different basis states. Figure \ref{fig:Dephasing} shows the outcome of Ramsey spectroscopy in the five different bases defined in \ref{tab:disorder} . 

\begin{table}[h]
    \begin{tabular}{ p{2cm} p{3.7cm} p{3.5cm} p{2cm} p{2.5cm} }
    \toprule
    Definition & Wavefunction & Precession rate & Noise & $1/ T_2^* $ \\
    \hline
    $|\psi_1 \rangle$ & $(|0\rangle + |1\rangle)/\sqrt{2}$ & $\gamma_{NV} B_{\parallel}$ + $d_{\parallel}E_{\parallel}$ & $\delta B_\parallel, \delta E_\parallel$ & $\pi [\Gamma_{B_{\parallel}} + \Gamma_{E_{\parallel}}]$  \\ 
    $|\psi_2 \rangle$ & $(|0\rangle + |-1\rangle)/\sqrt{2}$ &$\gamma_{NV} B_{\parallel}$ + $d_{\parallel}E_{\parallel}$ & $\delta B_\parallel, \delta E_\parallel$ & $\pi [\Gamma_{B_{\parallel}} + \Gamma_{E_{\parallel}}]$\\
    $|\psi_3 \rangle$ & $(|1\rangle + |-1\rangle)/\sqrt{2}$ & 2$\gamma_{NV} B_{\parallel}$ & $\delta B_\parallel$ & $2\pi \Gamma_{B_{\parallel}}$\\ 
    $|\psi_4 \rangle$ & $(|0\rangle + |D\rangle)/\sqrt{2}$ & $d_{\parallel}E_{\parallel}$ + $d_{\bot}E_{\bot}$ & $\delta E_\parallel$, $\delta E_\bot$ & $\pi [\Gamma_{E_{\parallel}} + \Gamma_{E_{\bot}}]$ \\ 
    $|\psi_5 \rangle$ & $(|0\rangle + |B\rangle)/\sqrt{2}$ & $d_{\parallel}E_{\parallel}$ + $d_{\bot}E_{\bot}$ & $\delta E_\parallel$, $\delta E_\bot$ & $\pi [\Gamma_{E_{\parallel}} + \Gamma_{E_{\bot}}]$ \\ 
    \hline
    \end{tabular}
    \caption{\label{tab:disorder} Five different basis states used for characterizing the local on-site disorder. The dark ($|D\rangle \equiv (|1\rangle - |-1\rangle)/\sqrt{2}$ ) and bright states ( $|B\rangle \equiv (|1\rangle + |-1\rangle)/\sqrt{2}$) are prepared by applying an off-axis magnetic field perpendicular to an NV symmetry axis. $\Gamma$ is a noise source-dependent inhomogeneous broadening contributing to the linewidth of the ESR. }
\end{table} 

As seen in the table \ref{tab:disorder}, each coherent superposition can effectively probe different types of noise components, enabling us to quantify the relative strengths of the on-site potential disorder. 
Using the identity $\Gamma = 1/\pi T_2^*$ and the relations given in the last column of table \ref{tab:disorder}, we can estimate a value for the different noise sources $\Gamma_{B_{\parallel}}$, $\Gamma_{E_{\parallel}}$, and $\Gamma_{E_{\bot}}$. 
The discrepancy in $T_2^*$ between $|\psi_1\rangle$ and $|\psi_2\rangle$ (as well as $|\psi_4\rangle$ and $|\psi_5\rangle$) in experimental data is presumably due to frequency-dependent field noise. 
By averaging these results, we can extract the three inhomogeneous broadening factors as $\Gamma_{B_{\parallel}}$ = 3.78(3) MHz, $\Gamma_{E_{\parallel}}$ = 2.18(8) MHz and $\Gamma_{E_{\bot}}$ = 4.30(13) MHz. 
The measured ESR linewidth $\Gamma_{\textrm{meas}} = \sqrt{8\ln2} W \approx$ 9.4 MHz (see Fig.~1D, main text) roughly agrees up to a factor of $\sim$1.5 with the calculated $\Gamma_{\textrm{calc}} \approx $ 6.0 MHz. 
According to this analysis, the random on-site disorder in our sample seems to result from both electric and magnetic fields with comparable weights. 

\subsection{Estimation of NV Density and Dipolar Interaction Strength}

\begin{figure}[h!]
\begin{center}
\includegraphics[width=0.4\textwidth]{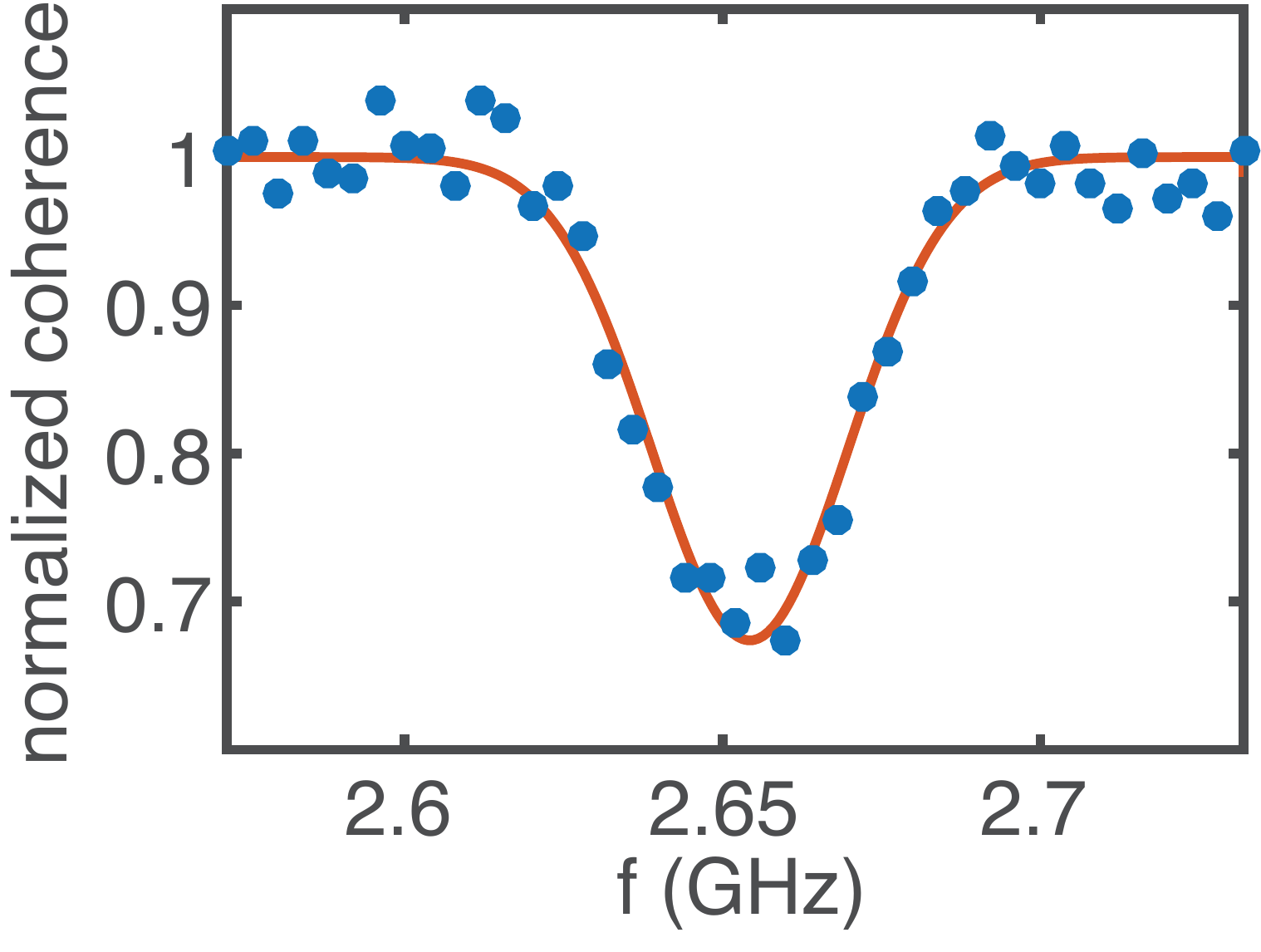}
\end{center}
\caption{\textbf{Intra-group Interaction Probed via Double Electron-Electron Resonance.} The relative, normalized spin echo coherence time at a fixed
time $\tau$ as a function of driving frequency of group B.}
\label{fig:deer}
\end{figure}

Due to the high density of NVs within our sample, the spin-echo coherence time is limited by interactions, as discussed in the main text. In particular, using the double electron-electron resonance (DEER) sequence presented in Fig. 2A in the main text, we verified experimentally that the additional dephasing of group A indeed originates from interactions with group B. Fig.~\ref{fig:deer} shows a measurement result of the DEER sequence in which we probe the relative spin-echo amplitude at a fixed time $\tau$ as a function of driving frequency of group B. It shows a clear resonance when $\omega = \omega_0^B$, indicating that inter-group interactions between group A and B lead to enhanced dephasing.

To quantitatively analyze the dependence of decoherence rate on the spin density, we study the dynamics of interacting spins using the exact diagonalization method with the effective Hamiltonian of Eq. (\ref{eq:Ham_A}). Comparing the numerical result to the experimental data  allows us to extract the density of NV spins in our sample. 

\begin{figure}[h!]
\begin{center}
\includegraphics[width=0.9\textwidth]{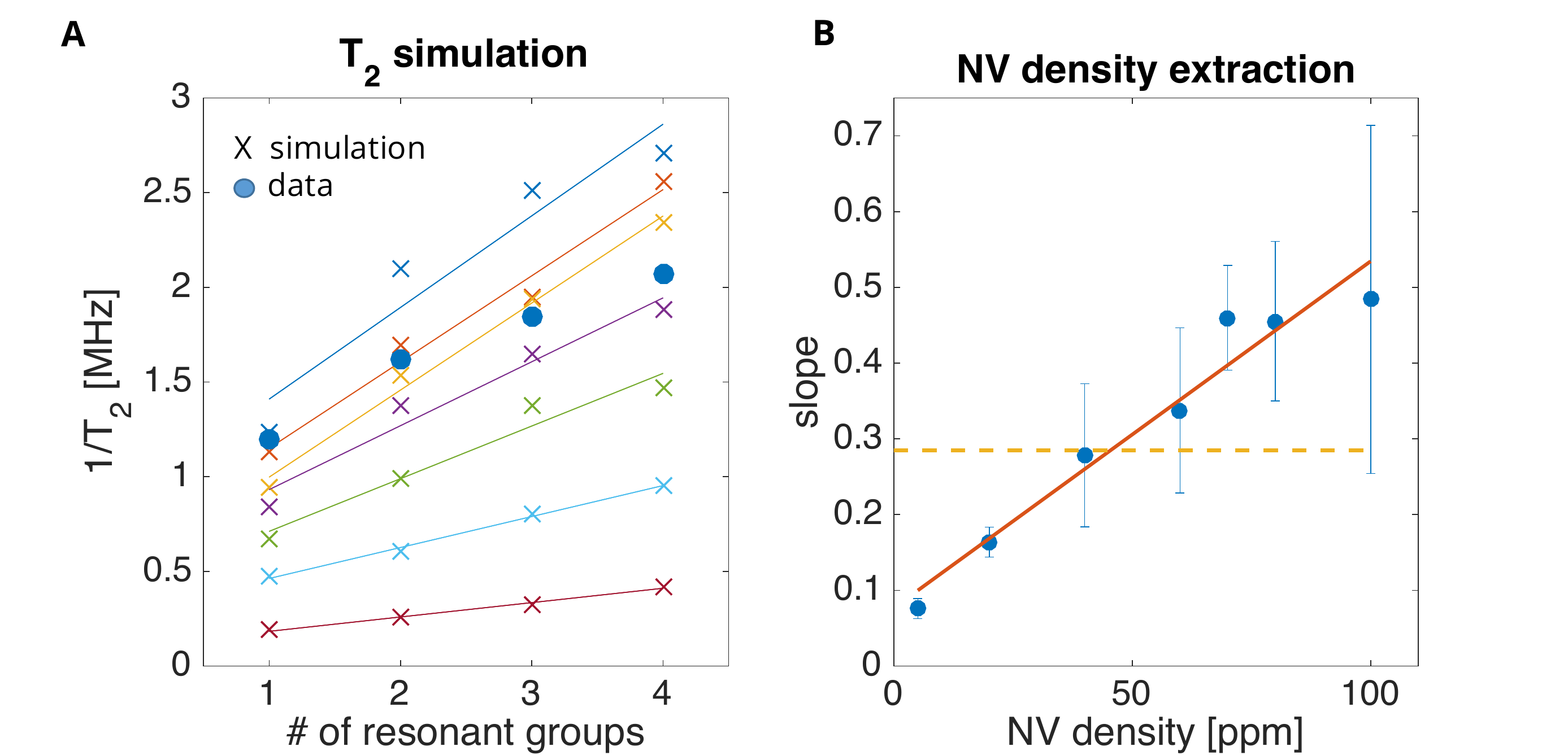}
\end{center}
\caption{\textbf{NV Density Extraction via Spin Echo Simulation.} \textbf{(A)} Comparison of the spin echo simulation results at different concentrations (crosses) to the measured data (circles). The total NV concentrations selected for the simulation are 5, 20, 40, 60, 70, 80 and 100 ppm. Solid lines are linear fits to the simulation to extract both $\gamma_b$ and $\gamma_{0}$ in the main text. \textbf{(B)} The NV concentration can be extracted by comparing the slopes ($\gamma_{0}$) taken from the numerical simulations to the extracted slope of the experiment data (orange dashed line).}
\label{fig:EDecho}
\end{figure}

Specifically, we simulate the time evolution of 12 NV spins under a spin echo pulse sequence protocol. The total NV concentrations selected for simulations are 5, 20, 40, 60, 70, 80 and 100 ppm. We averaged over $\sim$500 realizations of positional disorder, resulting in a single smooth coherence curve under the spin echo sequence. We fit the coherence decay with a simple exponential function and extract the decoherence rate, $\gamma_T \equiv 1/T_2$. 
Fig.~\ref{fig:EDecho}A summarizes the spin echo simulation results as a function of the number of resonant NV groups (effective density), where a linear dependence of $\gamma_T$ is identified for all the density values. We model the decoherence rate as $\gamma_T(\nu) = \gamma_b(\nu) + \nu \gamma_{0}(\nu)$, where $\nu$ is the number of resonant NV groups, $\gamma_b$ and $\gamma_{0}$ are density-dependent, bare and dipolar interaction-induced dephasing rates, respectively. Such linear dependence of $\gamma_T$ on $\nu$ is also confirmed in the experiment (see Fig. 2B in main text). By comparing $\gamma_{0}$ between the experiment and the simulation, we estimate the NV density in our sample to be $\sim$45 ppm (see Fig. \ref{fig:EDecho}B).

\subsection{Inhomogeneity of the Microwave Field}
\begin{figure}[h!]
\begin{center}
\includegraphics[width=0.5\textwidth]{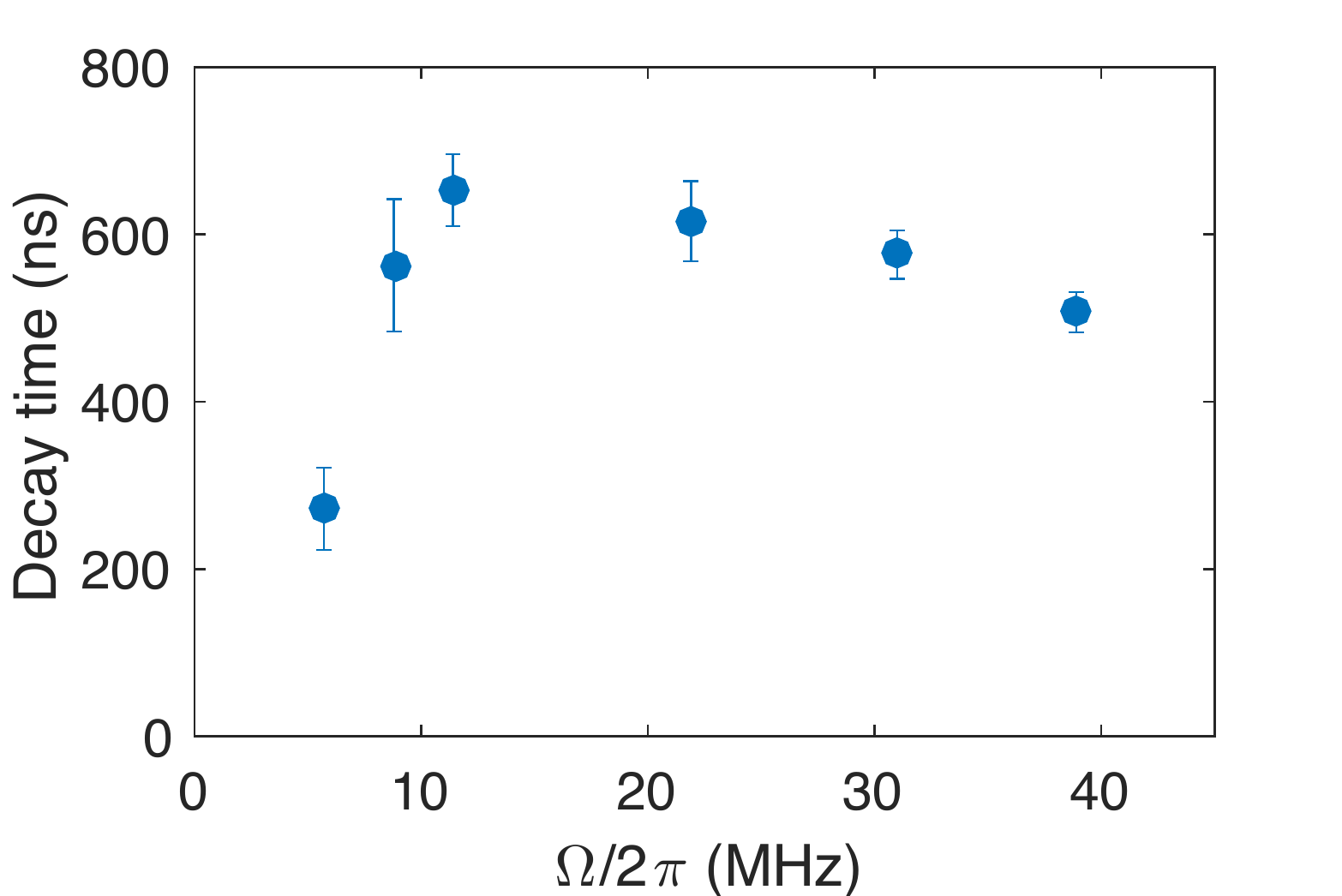}
\end{center}
\caption{\textbf{Rabi Oscillation Measurement.} Decay time of Rabi oscillations as a function of Rabi frequency $\Omega$.  }
\label{fig:RabiDecay}
\end{figure}

Hartmann-Hahn resonances rely on the exact matching of Rabi frequencies of two driving fields $\Omega_A = \Omega_B$.
Hence, stable and precise control of the driving strength is essential in our experiments.
To this end, we estimate the inhomogeneity of our microwave driving field, by measuring the decay time of Rabi oscillations at various driving strengths (Fig.~\ref{fig:RabiDecay}).

In an ideal case, the lifetime of Rabi oscialltions generally increases due to suppression of disorder (T$_2^*$). At higher driving strength (W$_\textrm{eff} \sim \delta^{I}$) this lifetime should saturate due to the effect of Ising interaction.
In our measurements however we observe a slight decrease in lifetime at high driving strengths, which is well explained by a 1.1$\%$ variation in Rabi frequency.
We attribute this variation to spatial inhomogeneity in the driving field.
With the strongest driving in our measurement $\Omega =(2\pi)$~32~MHz, this effect leads to a spread in Rabi frequencies of $\sim(2\pi)$~0.3~kHz. While it is still smaller than the effective disorder $\sim(2\pi)$~0.6~kHz, such an inhomogeniety ultimately limits the maximum driving strength of our thermalization experiments.

\section{Effective Hamiltonian of a Driven System}

In this section, we derive the effective Hamiltonian for a driven, dipolar interacting spin ensemble. The main idea is to work in a frame that is rotating along each NV group's quantization axis at corresponding driving frequency ($\omega_0^A$ and $\omega_0^B$ for group A and B, respectively).
If the difference between $\omega_0^A$ and $\omega_0^B$ is large compared to the interaction strength, then one can ignore exchange interactions between spins from different groups (secular approximation). This results in distinct forms of intra- and inter-group interactions.
We project the original Hamiltonian into two-level systems, and derive the effective Hamiltonian.

We start with the Hamiltonian for dipolar interacting NV centers
\begin{align}
    H = \sum_i H^0_i + \sum_i H^d_i(t) + \sum_{ij} H^{dd}_{ij},
\end{align}
where $H^0_i$ is a single particle Hamiltonian for a spin at site $i$, $H^{d}_i(t)$ is time-dependent driving, and $H^{dd}_{ij}$ is the magnetic dipole-dipole interaction between spins at sites $i$ and $j$.
The first term $H^0_i$ includes Zeeman coupling to an external magnetic field, the zero field splitting of a NV center, and any other disordered potentials arising from couplings to paramagnetic impurities as described in the main text.
In our experiments, dominant contributions for $H_i^0$ come from the zero-field splitting $\sim (2\pi)~2.87$GHz and Zeeman field projected along the quantization axis (a few hundred MHz), which are two orders of magnitude larger than the rest of the couplings. Setting $\hbar = 1$, we can write
\begin{align}
    \label{eqn:orientation_dep_ham}
    H^0_i \approx  (\Delta_0+\delta_{0,i}) \left(\hat{c}_i \cdot \vec{S}_i \right)^2 + (\Delta_B(\hat{c}_i) + \delta_{B,i}) \left(\hat{c}_i \cdot \vec{S}_i\right)
\end{align}
where $\vec{S}_i$ are spin-1 vector operators, $\hat{c}_i$ is the unit vector along the quantization axis of the spin, $\Delta_0 = (2\pi)~2.87$~GHz is the zero-field splitting, $\Delta_B(\hat{c}_i)$ is the Zeeman splitting along $\hat{c}_i$, and $\delta_{0,i}$ and $\delta_{B,i}$ are on-site disorder potentials. 
If the external magnetic field $\vec{B}$ is oriented in a way that $\Delta_B(\hat{c}_i)$ for different groups are sufficiently separated (compared to the driving strength), one can effectively address distinct groups independently. Below we assume such a case and consider resonant driving of two groups $A$ and $B$ using microwave frequencies $\omega_0^{A(B)} = \Delta_0 - \Delta_B(\hat{c}_{A(B)})$.
The Hamiltonian for such driving is given as $H^d_i(t) = \gamma_{NV}  \vec{B}_\textrm{MW} \cdot \vec{S}_i \cos\left( \omega_0 t \right)$, where $\gamma_{NV}$ is the gyromagnetic ratio of the NV center, and $\vec{B}_\textrm{MW}$ is the microwave field vector.
Now moving into the rotating frame with unitary transformation $U(t) = \exp{\left[-i \left( \sum_i \Delta_0 (\hat{c}_i \cdot \vec{S}_i)^2 + \Delta_B(\hat{c}_i) (\hat{c}_i \cdot \vec{S}_i)\right) t \right]}$ and applying rotating wave approximations, we obtain the effective single particle Hamiltonian
\begin{align}
    \bar{H}_i &= U^\dagger (t) \left[H_i^0 + H_i^d(t)\right] U(t) -i U^\dagger \frac{d}{dt} U
    \\ &= \left(\delta_{0,i} + \delta_{B,i}\right) \ketbra{1}{1} + \left(\delta_{0,i} - \delta_{B,i}\right)\ketbra{-1}{-1} + \frac{\Omega}{2}\left( \ketbra{-1}{0} + h.c. \right),
\end{align}
where $\{\ket{1}, \ket{0}, \ket{-1}\} $ is the basis of spin states along its quantization axis and $\Omega$ is the Rabi frequency of the driving.

The effective interaction among spins can be obtained in a similar way as follows. We start with the dipole-dipole interaction between spin-$i$ and spin-$j$
\begin{align}
\label{eqn:angle_dep_dipolar}
    H^{dd}_{ij} =  -\frac{J_0}{r^3}
    \left(
    3 \left(\vec{S}_i \cdot \hat{r}\right)
    \left(\vec{S}_j \cdot \hat{r}\right)
    -
    \vec{S}_i \cdot \vec{S}_j
    \right),
\end{align}
where $J_0 = (2\pi)~52~\text{MHz}\cdot\text{nm}^3$ and $\vec{r}$ is the relative position between two spins.
In the rotating frame, we obtain the effective interaction by replacing $\vec{S}_i \mapsto U^\dagger(t) \vec{S}_i U(t)$.
Since we are interested in the interaction in the basis of each NV's own quantization axis, we first explicitly rewrite $\vec{S}_i$ in terms of $(S_i^x, S_i^y, S_i^z)$ in a coordinate system where $\hat{z}_i$ is parallel to the quantization axis $\hat{c}_i$
\begin{align}
    \bar{H}^{dd}_{ij} = U^\dagger (t) H^{dd}_{ij}U(t) = -J_0/r^3
    \Big [
        &
        \big( 
            3\left(\hat{r}\cdot \hat{x}_i\right)\left(\hat{r}\cdot \hat{x}_j\right)
            -
            \hat{x}_i \cdot\hat{x}_j
        \big) S^x_i S^x_j\\
        &
        +\big( 
            3\left(\hat{r}\cdot \hat{y}_i\right)\left(\hat{r}\cdot \hat{y}_j\right)
            -
            \hat{y}_i \cdot\hat{y}_j
        \big) S^y_i S^y_j\\
        &
        +\big( 
            3\left(\hat{r}\cdot \hat{x}_i\right)\left(\hat{r}\cdot \hat{y}_j\right)
            -
            \hat{x}_i \cdot\hat{y}_j
        \big) S^x_j S^y_j\\
        &
        +\big( 
            3\left(\hat{r}\cdot \hat{y}_i\right)\left(\hat{r}\cdot \hat{x}_j\right)
            -
            \hat{y}_i \cdot\hat{x}_j
        \big) S^y_i S^x_j\\
        &
        +\big( 
            3\left(\hat{r}\cdot \hat{z}_i\right)\left(\hat{r}\cdot \hat{z}_j\right)
            -
            \hat{z}_i \cdot\hat{z}_j
        \big) S^z_i S^z_j
    \Big ]\\
    +
     H_\textrm{rest},
\end{align}
where $H_\textrm{rest}$ contains all the other terms of the form $S^xS^z, S^yS^z, S^zS^x, S^zS^y$.

We now perform rotating wave approximations. This is very well justified because the typical strength of the interaction is much weaker than the driving frequency $J_0/r^3 \sim (2\pi)~0.4$~MHz $\ll \omega_0^{A,B} \sim (2\pi)~2.5$~GHz. 
First, we note that $S^x$ and $S^y$ operators are rapidly oscillating in time while $S^z$ remains invariant, $[S^z_i, U(t)] = 0$. Therefore, every term in $H_\textrm{rest}$ may be safely ignored.
Then, introducing  
\begin{align}
    g^{+}_{ij} =& \frac{1}{2}
    \Big[ 
            3\left(\hat{r}\cdot \hat{x}_i\right)\left(\hat{r}\cdot \hat{x}_j\right)
            -
            \hat{x}_i \cdot\hat{x}_j
            +
             3\left(\hat{r}\cdot \hat{y}_i\right)\left(\hat{r}\cdot \hat{y}_j\right)
            -
            \hat{y}_i \cdot\hat{y}_j
    \Big]
     \\
    g^{-}_{ij} =& \frac{1}{2}
        \Big[ 
            3\left(\hat{r}\cdot \hat{x}_i\right)\left(\hat{r}\cdot \hat{x}_j\right)
            -
            \hat{x}_i \cdot\hat{x}_j
            -
             3\left(\hat{r}\cdot \hat{y}_i\right)\left(\hat{r}\cdot \hat{y}_j\right)
            +
            \hat{y}_i \cdot\hat{y}_j
    \Big]
    \\
    h^{+}_{ij} =& \frac{1}{2}
        \Big[
            3\left(\hat{r}\cdot \hat{x}_i\right)\left(\hat{r}\cdot \hat{y}_j\right)
            -
            \hat{x}_i \cdot\hat{y}_j
        +
            3\left(\hat{r}\cdot \hat{y}_i\right)\left(\hat{r}\cdot \hat{x}_j\right)
            -
            \hat{y}_i \cdot\hat{x}_j
        \Big]
             \\
    h^{-}_{ij} =& \frac{1}{2}
        \Big[
            3\left(\hat{r}\cdot \hat{x}_i\right)\left(\hat{r}\cdot \hat{y}_j\right)
            -
            \hat{x}_i \cdot\hat{y}_j
        -
            3\left(\hat{r}\cdot \hat{y}_i\right)\left(\hat{r}\cdot \hat{x}_j\right)
            +
            \hat{y}_i \cdot\hat{x}_j
        \Big]\\
        q_{ij} = &             3\left(\hat{r}\cdot \hat{z}_i\right)\left(\hat{r}\cdot \hat{z}_j\right)
            -
            \hat{z}_i \cdot\hat{z}_j,
\end{align}
we can simply rewrite
\begin{align}
\label{eqn:dd_int_rot_1}
    \bar{H}^{dd}_{ij} \approx -J_0/r^3
    \big[
        & g^+_{ij} (S^x_i S^x_j + S^y_i S^y_j)
        + h^-_{ij} (S^x_i S^y_j - S^y_i S^x_j) 
        + q_{ij} S^z_i S^z_j \\
  \label{eqn:dd_int_rot_11}
        &+ g^-_{ij} (S^x_i S^x_j - S^y_i S^y_j)
        + h^+_{ij} (S^x_i S^y_j + S^y_i S^x_j)
    \big].
\end{align}
Here, $g^+$ and $h^-$ terms correspond to ``flip-flop'' type transitions, exchanging one unit of spin polarization,
\begin{align}
    (S^x_i S^x_j + S^y_i S^y_j) 
    =&
        \ketbra{+0}{0+}
        + \ketbra{+-}{00}
        + \ketbra{00}{-+}
        + \ketbra{0-}{-0}
        + h.c.\\
    (S^x_i S^y_j - S^y_i S^x_j) 
=& i \big(
       \ketbra{+0}{0+}
        + \ketbra{+-}{00}
        + \ketbra{00}{-+}
        + \ketbra{0-}{-0} \big)
        + h.c.
\end{align}
In addition, owing to the strong anharmonic level structure, we may also ignore flip-flop transitions between levels with large energy differences, e.g. terms such as $\ketbra{+-}{00}$.
Finally, we ignore the terms in Eq.~\eqref{eqn:dd_int_rot_11} as they correspond to double flip-up or flip-down and rapidly oscillate in time. After these approximations, the effective interaction becomes
\begin{align}
\label{eqn:secular_dipole_interaction}
    \bar{H}^{dd}_{ij}\approx
    - J_0 /r^3 
    \big[ 
            \left( g^+_{ij} + i h^-_{ij}\right) \ketbra{+0}{0+}
        + \ketbra{0-}{-0}
        + h.c.
        + q_{ij} S^z_i S^z_j
    \big].        
\end{align}
Now we divide into two cases depending on whether spins $i$ and $j$ belong to the same group or to different groups.
In the former case, the quantization axes coincide, and we can simplify $h^-_{ij} = 0$, $g^+_{ij} = \frac{1}{2} (1- 3\cos^2 \theta)$, and $q_{ij} = -(1-3\cos^2\theta)$ with $\cos \theta \equiv \hat{z} \cdot \hat{r}$. In the latter case, the flip-flop terms are again rapidly oscillating, and only the Ising interaction $S_i^z S_j^z$ remains, resulting in
\begin{align}
\label{eqn:eff_interaction_01}
\bar{H}^{dd}_{ij} \approx
\left\{ 
\begin{array}{cc}
- \frac{J_0q_{ij}}{r^3}  \left(- \frac{\ketbra{+0}{0+}
        + \ketbra{0-}{-0}
        + h.c.
}{2} +  S_i^z S_j^z\right) & \textrm{same group}\\
-\frac{J_0q_{ij} }{r^3} S_i^z S_j^z & \textrm{different groups}
\end{array}
\right.
.
\end{align}
These interactions as well as the single particle terms conserve the total population of spins in $\ket{+}$.
Therefore, once the system is initialized into a state with no population in $\ket{+}$, the dynamics remains in the manifold spanned by $\ket{-}$ and $\ket{0}$. Projecting $\sum_i \bar{H}_i + \sum_{ij} \bar{H}^{dd}_{ij} $ into this manifold, we obtain the Hamiltonian for an effective two-level system.
Introducing spin-1/2 operators $\vec{s}$ for two levels $\ket{-}$ and $\ket{0}$, we obtain $H_T = H_A + H_B + H_{AB}$, where
\begin{align}
H_{A (B)} &= \sum_{i\in A(B)} [(\delta_{0,i} - \delta_{B,i}) s_i^z + \Omega_{A(B)} s_i^x] + \sum_{i,j\in A(B)}\frac{J_0 q_{ij}}{r_{ij}^3} \left( s_i^x s_j^x + s^y_i s_j^y - s^z_i s^z_j\right), \label{eq:Ham_A}\\
 H_{AB} &= - \sum_{i\in A,j\in B} \frac{J_0 q_{ij} }{r_{ij}^3} s_i^{z_A} s_j^{z_B},
\end{align}
up to a constant.

Finally, we remark one particularly interesting aspect of this Hamiltonian in the dressed-state basis, i.e., quantization along $s^x_i$.
With sufficiently strong driving, $s^x_i$ becomes a good spin polarization basis, and one can rewrite the interactions in terms of $s^{\pm} = s^y \pm i s^z$, wherein the intra-group interaction becomes $\propto s_i^x s_j^x + (s^+_i s^+_j + s^-_i s^-_j)/2$ and the inter-group interaction $\propto (s_i^+ s_j^- + s_i^+ s_j^+ + h.c. )$.
Here, we find that spin exchange terms ($s_i^+ s_j^- + h.c.$) are missing in the intra-group interaction.
Omitting the energy non-conserving terms such as $s_i^+ s_j^+$ (secular approximation with a strong driving strength $\Omega$), we obtain the effective Hamiltonian described in the main text.

\section{Resonance Counting Theory}
In this section, we provide a detailed study of the single particle resonance counting theory.
We will first focus on the case of quenched on-site potential disorder, deriving the disorder-dependent power-law relaxation presented in the main text.
Then, we generalize the result to the case when disordered potentials are time-dependent.

\subsection{Disorder-dependent Power-law Decay}

As discussed in the main text, we estimate the survival probability of a single spin excitation based on a simple counting argument.
At time $t$, we compute the probability $\textrm{Pr}(k;t)$ that the central spin is connected to $k-1$ other spins via a network of resonances, as defined in the main text.
Assuming that the population of the excitation is equally shared among a resonating cluster, the survival probability is given as
\begin{align}
\label{eqn:pksum}
P(t) \approx \sum_{k=1}^\infty \frac{1}{k} \textrm{Pr}(k;t).
\end{align}
reducing our problem to the computation of $\textrm{Pr}(k;t)$. Below we will show that the dominant contributions arise from $k=1$, suggesting that finding a single resonant partner is usually enough to delocalize the spin excitation over the entire sample. 

In general, the exact  calculation of $\textrm{Pr}(k;t)$ is difficult.
This is because the connectivity of the resonance network is correlated due to the spatial structure ($d$-dimensional Euclidean space) as well as a given assignment of random on-site potentials, e.g., if spin pairs $(a,b)$ and $(b,c)$ are pair-wise resonant, it is likely that the pair $(a,c)$ is also resonant, etc.
However, the qualitative behavior of $\textrm{Pr}(k;t)$ can still be well-understood by ignoring these correlations. 
In such a case, we may assume that the number of resonant partners $\ell$ for a spins is drawn from a probability distribution $p(\ell)$ and that this process can be iterated for each partner.
We note that such a process may not terminate,
in which case the central excitation becomes delocalized over a macroscopic number of spins.
We first compute $p(\ell)$ as a function of time $t$.
For $\ell=0$, a spin of interest (spin-$i$) must not have any resonating spins at any distance from $r_\textrm{min}$ to $R(t) \equiv (J_0 t)^{1/3}$, where $r_\textrm{min}$ is the short-distance cut-off.
Hence, $p(0;t)$ is given as a product of probabilities:
\begin{align}
    p(0;t) &= \prod_{r_\textrm{min} \leq r < R(t)} 
    \left(
        1 - 4\pi n r^{2} dr \frac{ \beta J_0/r^3}{W_\textrm{eff} } \right)\\
        \label{eqn:secular_dipole_interaction}
    &= \exp{
    \left[
    - \int_{r_\textrm{min}}^{R(t)}    
        \frac{4\pi n Q_{\textrm{res}}}{r} dr    
    \right]
    }
\end{align}
where $4\pi n r^{2}dr$ is the probability of finding a spin at distance $r$, and $Q_{\textrm{res}} = \beta J_0/( W_\textrm{eff} r^3)$ is the probability that the spin resonates with the spin-$i$.
Defining $\lambda(t) = 4\pi Q_{\textrm{res}} (\ln{R(t)} - \ln{r_\textrm{min}})$,  we obtain $p(0;t) = \exp{\left[-\lambda(t)\right]} $.
Similarly, we can calculate $p(\ell;t)$ for $\ell > 0$, and obtain   
$  p(\ell;t) = \frac{1}{\ell!}  \left(\lambda(t)\right)^\ell e^{-\lambda(t)}$, which is the Poisson distribution with mean $\lambda(t)$.

To show that the dominant contribution of Eq.~\eqref{eqn:pksum} arises from the $k=1$ term, we consider the probability of the termination of the resonance finding process, $P_\textrm{term}$. It satisfies the self-consistency equation
\begin{align}
\label{eqn:termination_prob}
P_\textrm{term} = e^{-\lambda} + \sum_{\ell=1}^\infty \frac{\lambda^\ell e^{-\lambda}}{\ell !}  \left(P_\textrm{term}\right)^\ell,
\end{align}
where the first term corresponds to the case where the initial spin does not have any resonance up to time $t$, while the second term implies the termination of each sub-graph generated from $\ell$ resonant spins.
For sufficiently large $\lambda$, $P_\textrm{term}$ becomes small, and its contribution is dominated by the first term ($\ell = 0$). In our case, $\lambda(t)$ is a function of time which diverges in the limit $t \rightarrow \infty$.
As we are interested in the late time dynamics, we may consider the first term only.
In terms of $\textrm{Pr}(k;t)$, this corresponds to approximating $\textrm{Pr}(k;t) \sim 0$ for $k > 2$.
Finally, noting that that $\textrm{Pr}(k=1;t) = p(0;t)$, we recover the expression in the main text. 

\begin{figure}[h!]
\begin{center}
\includegraphics[width=.4\textwidth]{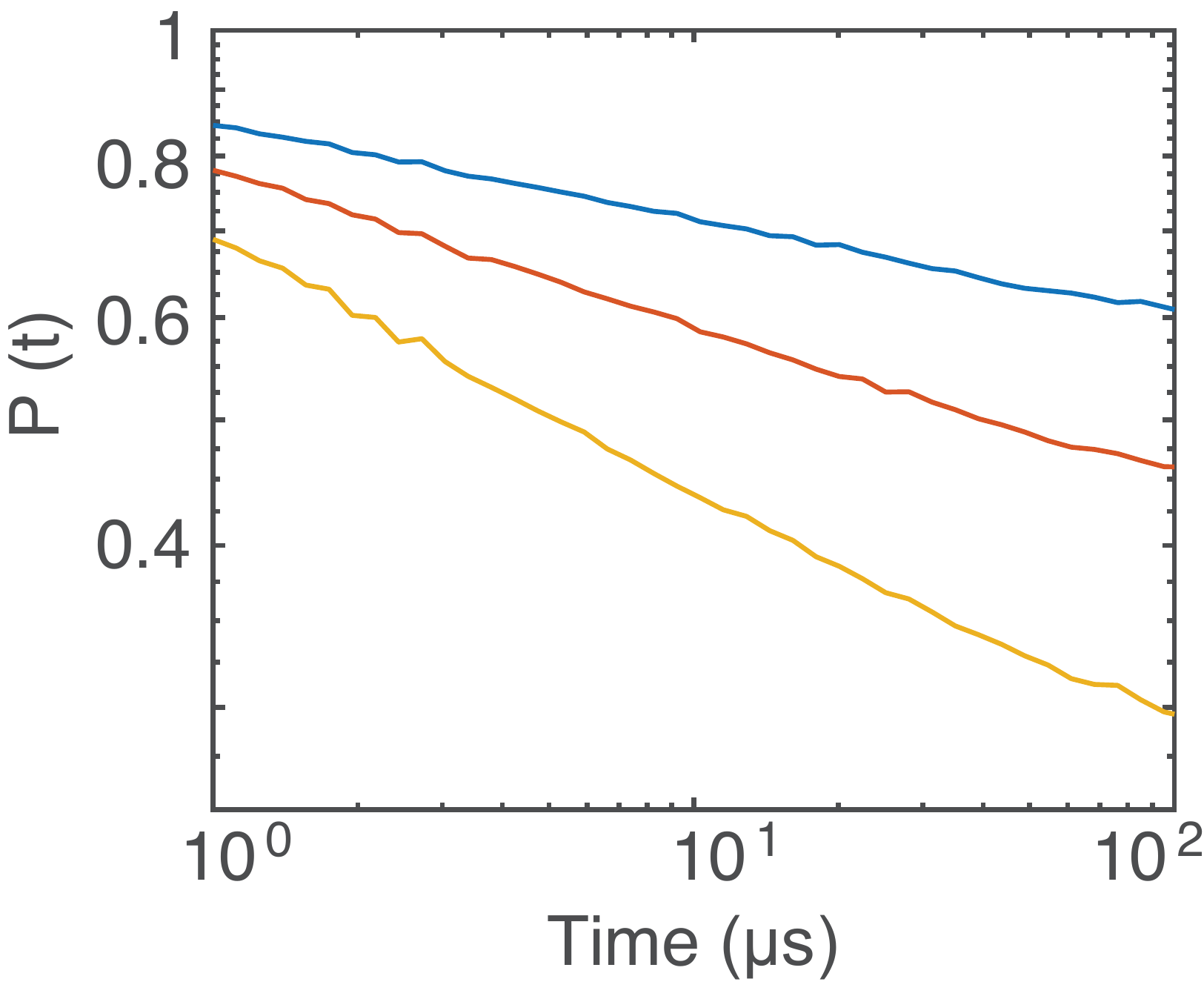}
\end{center}
\caption{\textbf{Single-particle simulation of power-law dynamics.}  Blue, red, and yellow curve correspond to $\Omega = (2\pi)$ 3, 8, and 20 MHz, respectively. For the simulations, we use $10^4$ spins and average over more than $10^3$ disorder realizations.}
\label{fig:powerlaw}
\end{figure}

We numerically test the analytic resonance counting that predicts the power-law decay dynamics. In the limit of single-particle excitation, the survival probability $P(t) = |\langle \psi(t) | \psi(0) \rangle|^2$ can be computed at any time $t$ after the time evolution of a system under $H_{\textrm{eff}}$ (See Eq. (1) in the main text). Considering physically relevant parameters used in the experiments, we verify such power-law decay dynamics for up to $10^4$ spins as shown in Fig.~\ref{fig:powerlaw}. Moreover, we confirm the extracted power-law exponent is inversely proportional to effective disorder $W_{\textrm{eff}}$ (Fig.~4C in the main text), further substantiating the thermalization mechanism based on rare resonances. The power-law exponents extracted from the simulations are summarized in Fig.~\ref{fig:HHfitparam}A.

\subsection{Interplay between Dimensionality and long-range Interaction}
The critical nature of a disordered dipolar spin ensemble in three dimensions originates from the interplay between long-range interactions and dimensionality. To see this, we can generalize the resonance counting analysis for a situation in which a single particle excitation is located in a $d$-dimensional spin system with long-range coupling decaying as $1/r^\alpha$. In such a setting, the survival probability $P(t)$ can be expressed as, 
\begin{align}
P(t) &= \exp{
    \left[
    - \int_{r_\textrm{min}}^{R(t)}    
        n S_d r^{d-1} \frac{\beta J_0/r^\alpha}{W_\textrm{eff}}dr   
    \right]} \\
    &= \exp{\left[ - \frac{n S_d \beta J_0}{W_\textrm{eff}} \int_{r_\textrm{min}}^{R(t)}  r^{d-\alpha-1} dr  \right]},
\end{align}
where $S_d$ is the surface area of the $d$-dimensional volume. In fact, the argument of $P(t)$, $\int_{r_\textrm{min}}^{R(t)} r^{d-\alpha-1} dr$, is associated with the probability of finding a resonance up to the distance $R(t)$ reachable at time $t$. Hence, when the dimensionality $d$ is larger (smaller) than the interaction strength $\alpha$, the above integral diverges (converges) as $R(t)$ becomes large, which implies delocalization (localization) of the single particle excitation. In the critical case where $d$ is equal to $\alpha$, the resonance probability increases at a slow logarithmic rate, resulting in the power-law relaxation of the initial spin state as derived in Eq. (2) in the main text. In the limit of single particle excitations we therefore associate our system dynamics to such criticality behavior. However, due to the presence of many-spin excitations, much richer dynamics may appear at longer times. We attribute the deviation of power-law dynamics at late times observed in our experiments to this effect. 

\subsection{Time-dependent Disorder}
Now we consider the case of time-dependent disorder.
For concreteness, we assume that the on-site potential disorder is given as a sum of a static and a dynamical disorder potential, $\tilde{\delta}_i (t) = \tilde{\delta}_i^s + \tilde{\delta}_i^d (t)$, where the static part $\tilde{\delta}_i^s$ (dynamical part $\tilde{\delta}_i^d (t)$) is random with zero mean and standard deviation $W_s$ ($W_d$). While $\tilde{\delta}_i^s$ is time-independent, the dynamical component $\tilde{\delta}_i^d (t)$ changes over time by uncorrelated jumps at a rate $\Gamma = 1/\tau_d$. Here, we focus on an experimentally relevant regime where $W_s \gg W_d \gtrsim nJ_0 >  1/\tau_d$.

As already mentioned in the main text, we modify the resonance criteria as follows. Two spins at sites $i$ and $j$ are on resonance at time $t$ if: (1) \emph{at any point in time} $t' < t$, their energy mismatch is smaller than their dipolar interaction strength, $|\tilde{\delta}_i (t') - \tilde{\delta}_j(t')| < \beta J_{ij}/r_{ij}^3$, and (2) the interaction occurs within the time-scale $t$, $J_{ij}/r_{ij}^3 > 1/t$.
While the second part of the condition is unchanged, the first part now captures that a pair may be brought into resonance by spectral jumps.
Under the hierarchy of $W_s \gg W_d \gtrsim nJ_0 > 1/\tau_d$, the condition (1) can be approximated by two independent events: (a) the static energy mismatch is small enough, $|\tilde{\delta}_i^s - \tilde{\delta}_j^s | < W_d$, and (b) the dynamical energy mismatch is smaller than the coupling strength, $ |\tilde{\delta}_i^d (t') - \tilde{\delta}_j^d (t') | < \beta J_{ij} / r_{ij}^3$ at some time $t' < t$.
In combination, the condition (1) is satisfied with the probability 
\begin{align}
\label{eqn:Pres}
P_\textrm{res}(r,t) 
\approx \frac{W_d}{W_s} \left( 1- e^{-\frac{\beta J_0/r^3}{W_d} \frac{t}{\tau_d}}\left( 1- \frac{\beta J_0/r^3}{W_d}\right) \right)
\end{align}
which is the product of probabilities for conditions (a) and (b).
For the second factor, we used the probability that the initial configuration is off-resonant, $(1 - \frac{\beta J_0/r^3}{W_d})$, and the probability that none of the subsequent spectral jumps brings them into resonance $e^{-\frac{\beta J_0/r^3}{W_d} \frac{t}{\tau_d}}$.
We note that, in practice, one should use $\textrm{max}(0, 1- \beta J_0/W_d r^3)$ instead of $(1- \beta J_0/ W_d r^3)$ since a probability cannot be less than zero.
Finally, the survival probability is obtained by requiring no resonance at every distance $r$ up to $R(t) = (J_0 t)^{1/3}$
\begin{align}
\label{eqn:tddisorder_2_surv_prob}
\overline{P(t)} = \exp{\left[ -\int_{r=r_0}^{R(t)} 4 \pi n r^2 P_\textrm{res}(r,t) dr \right]},
\end{align}
where $r_0$ is the short distance cut-off of the NV separations.
We use the cut-off distance  $r_0\sim1.4$~nm, at which the corresponding dipole-dipole interaction is $J_0/r_0^3 \sim (2\pi)~20$~MHz.
Due to dipole blockade, a pair of NV centers closer than $r_0$ cannot be addressed by microwave driving of Rabi frequency $\Omega\sim(2\pi)~20$~MHz, which we use for initial preparations of spin states. Those spins do not participate in the spin exchange dynamics due to large energy mismatch.
We note that, $\lim_{\Gamma \rightarrow 0} P_{\textrm{res}}(r,t) \rightarrow Q_{\textrm{res}}(r) = \beta J_0/( W_s r^3) $ and the Eq.~\eqref{eqn:tddisorder_2_surv_prob} correctly reduces to the disorder-dependent power-law decay. 
In the presence of a small but finite $\Gamma = 1/\tau_d$, integrating Eq.~\eqref{eqn:tddisorder_2_surv_prob} using Eq.~\eqref{eqn:Pres} yields,
\begin{align}
\overline{P(t)} = P_1(t) P_0(t),
\end{align}
where
\begin{align}
P_1(t) &= \exp\left[-\frac{4\pi n}{3}\frac{W_d}{W_s} \left\{ J_0(t - t_0) - J_0(t e^{-\frac{\beta}{W_d\tau_d}} - t_0 e^{-\frac{t}{t_0}\frac{\beta}{W_d\tau_d}} ) \right\} \right] \\
P_0(t) &= \exp\left[-\frac{4\pi nJ_0 \beta}{3W_s} \left\{(1+t/\tau_d) \mathcal{G}[0, \frac{\beta}{W_d\tau_d}] - (1+t/\tau_d) \mathcal{G}[0,\frac{t}{t_0}\frac{\beta}{W_d\tau_d}]  \right\} \right].
\end{align}
Here $\mathcal{G}$ is an incomplete Gamma function. In the limit of the hierarchy $W_s \gg W_d \gtrsim nJ_0 >  1/\tau_d$, we can simplify:
\begin{align}
P_1(t) &\approx C_1 \exp\left[-\frac{4\pi n}{3}\frac{W_d}{W_s} \left( J_0 t (1 - e^{-\frac{\beta}{W_d\tau_d}}) \right) \right] \\
&\approx C_1 \exp\left[-\frac{4\pi n J_0 \beta}{3 W_s}\frac{t}{ \tau_d} \right] \\
& \equiv C_1 \exp[-t/T^*],
\end{align}
where $C_1$ is a time-independent prefactor and 
\begin{align}
T^* = \frac{3W_s\tau_d}{4\pi nJ_0 \beta} \propto \frac{W_s \tau_d}{nJ_0}. 
\end{align}
Similarly, $P_0(t)$ can also be simplified as, 
\begin{align}
P_0(t) &\approx  C_2 \exp\left[-\frac{4\pi nJ_0 \beta}{3W_s} \left\{ \mathcal{G}[0,\frac{\beta}{W_d\tau_d}] - \mathcal{G}[0,\frac{t}{t_0}\frac{\beta}{W_d\tau_d}] \right\} \right] \\
&\approx C_2' \exp\left[-\frac{4\pi nJ_0 \beta}{3W_s} \ln(t/t_0) \right] \\
\label{eqn:P0}
&=  C_2' (t/t_0)^{-\frac{4\pi nJ_0 \beta}{3W_s}}.
\end{align}
Here we used the approximation $\mathcal{G}(0,z) \approx -\ln(z) + \gamma + \mathcal{O}(z)$ for $z \ll 1$. 
Once again, we rediscover the power-law decay (Eq.~\eqref{eqn:P0}) predicted in the main text, but now only up to a finite time $T^*$:
\begin{align}
\label{eqn:time_dep_disorder}
\overline{P(t)} = P_1(t) P_0(t) \propto e^{-t/T^*} t^{-\frac{4\pi nJ_0 \beta}{3W_s}}.
\end{align}
Therefore, according to the Eq.~\eqref{eqn:time_dep_disorder}, the weak time-dependent disorder results in a multiplicative exponential correction to the power-law decay up to  $t<T^*$, beyond which the thermalization accelerates substantially. Furthermore, our theory model predicts that $T^*$ is linearly proportional to the static disorder strength $W_s$, which is consistent with our observations (See Fig. 3D and Fig. 4D in the main text).

\section{Detailed Analysis of Thermalization Experiments}

\subsection{Effective Disorder Control under Spin-locking Conditions}
To investigate the interplay between disorder and interaction experimentally, it is required to tune both disorder and interaction in a controlled way. In our experiments, we rely on a spin-locking technique in which both the energy spacing and the on-site disorder of a spin ensemble can be controlled in a continuous fashion. 

As discussed in the main text, spin-locking allows us to prepare spins in the dressed state basis. In the new basis, the energy eigenstates are $\ket{\pm} \approx (\ket{m_s = 0} \pm \ket{m_s = -1})/\sqrt{2}$ and are split by an effective Rabi frequency of a spin-lock field, $\Omega_{\textrm{eff}} = \sqrt{\Omega^2 + \delta^2}$, where $\Omega$ is the driving strength and $\delta$ is the on-site disorder in the bare frame. Owing to a random distribution of $\delta$, the new level spacing $\Omega_{\textrm{eff}}$ is also a random variable. Therefore, an effective disorder under the spin-locking condition can be defined as, 
\begin{align}
W_{\textrm{eff}} \equiv \sqrt{\textrm{Var}[\Omega_\textrm{eff}]} = \sqrt{E[\Omega^2 + \delta^2] - E[\sqrt{\Omega^2 + \delta^2}]^2},
\end{align}
where $\textrm{Var}[X]$ and $E[X]$ are the variance and expectation value of a random variable X. Since the disorder in the bare frame follows a Gaussian distribution with a standard deviation $W$,  the expectation values can be expressed as
\begin{align}
E[\Omega^2 + \delta^2] &= \frac{1}{\sqrt{2\pi W^2}}\int_{-\infty}^{+\infty} d\delta \; [\Omega^2 + \delta^2]  e^{-\delta^2/2W^2} \\
E[\sqrt{\Omega^2 + \delta^2}] &=  \frac{1}{\sqrt{2\pi W^2}}\int_{-\infty}^{+\infty} d\delta \; \sqrt{\Omega^2 + \delta^2}  e^{-\delta^2/2W^2}.
\end{align}
In the case of weak driving ($\Omega \ll \delta$), $W_\textrm{eff} \approx \sqrt{\textrm{Var}[\delta]} = W$; namely, the effective disorder is almost equal to that in the bare frame. However, as the driving strength $\Omega$ increases we find $W_\textrm{eff} \approx \sqrt{\textrm{Var}[\frac{\delta^2}{2\Omega}]} = \frac{W^2} {\sqrt{2}\Omega}$. Hence, the effective disorder $W_\textrm{eff}$ can be tuned by adjusting the Rabi frequency $\Omega$ in the dressed state basis.

\begin{figure}[h!]
\begin{center}
\includegraphics[width=.5\textwidth]{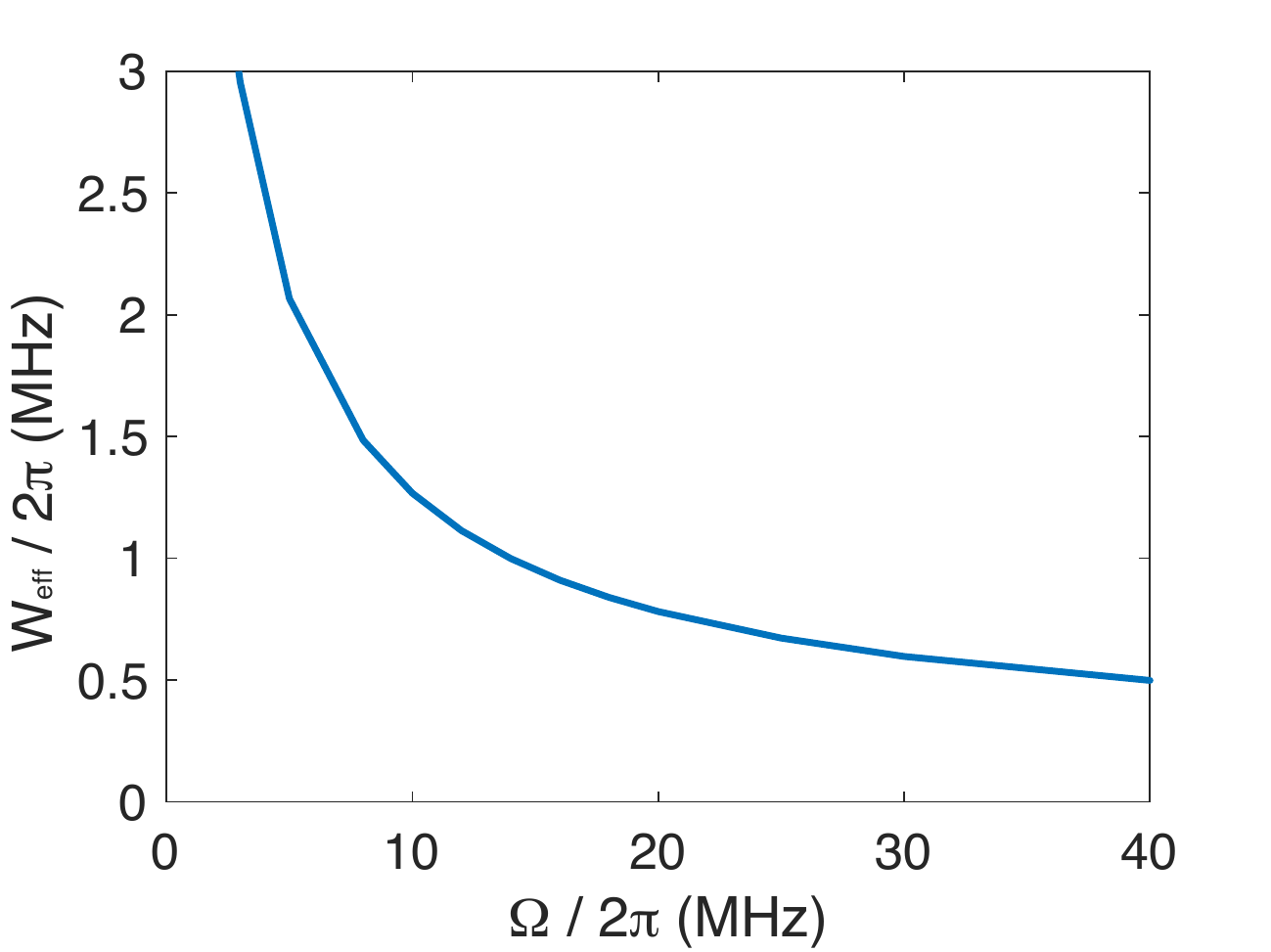}
\end{center}
\caption{\textbf{Effective disorder under spin locking conditions.}  Based on the resonance counting argument, the effective disorder $W_\textrm{eff}$ can be computed as a function of the Rabi frequency $\Omega$.}
\label{fig:effdisorder}
\end{figure}

We note that the probability distribution of $\Omega_\textrm{eff}$ is highly asymmetric, which may lead to small corrections to our counting argument at a quantitative level.
To this end, for our numerical computations, we use an alternative definition of $W_\textrm{eff}$ which is consistent with our resonance counting argument. Recall that two spins at site $i$ and $j$ with separation $r$ are defined to be on resonance 
when $|\tilde{\delta_i} - \tilde{\delta_j}| < \beta J_0/r^3$ and that we assumed this occurs with probability $Q_\textrm{res} \propto (J_0/r^3) / W_\textrm{eff} $.
Therefore, the effective disorder strength $W_\textrm{eff}$ should be defined in the same way from the full distribution of $\Omega_\textrm{eff}$.
More specifically, we compute the probability $q(\xi)$ that two independent random variables $\tilde{\delta}_{i}$ and $\tilde{\delta}_{i}$ satisfy $|\tilde{\delta}_i - \tilde{\delta}_j| < \xi$ for a small parameter $\xi$. In the limit of $\xi \ll W^2/\sqrt{2} \Omega$, the probability $q(\xi)$ is linearly proportional to $\xi$. Then, we define the effective disorder as $W_\textrm{eff} \equiv \lim_{\xi\rightarrow 0} \xi /q(\xi)$.
Fig.~\ref{fig:effdisorder} shows the dependence of $W_\textrm{eff}$ as a function of $\Omega$. 
In the limit of large $\Omega$, the effective disorder scales as $W_\textrm{eff} \propto 1/\Omega$, as expected.

\subsection{Effects of Incoherent Dynamics}
\begin{figure}[h!]
\begin{center}
\includegraphics[width=1\textwidth]{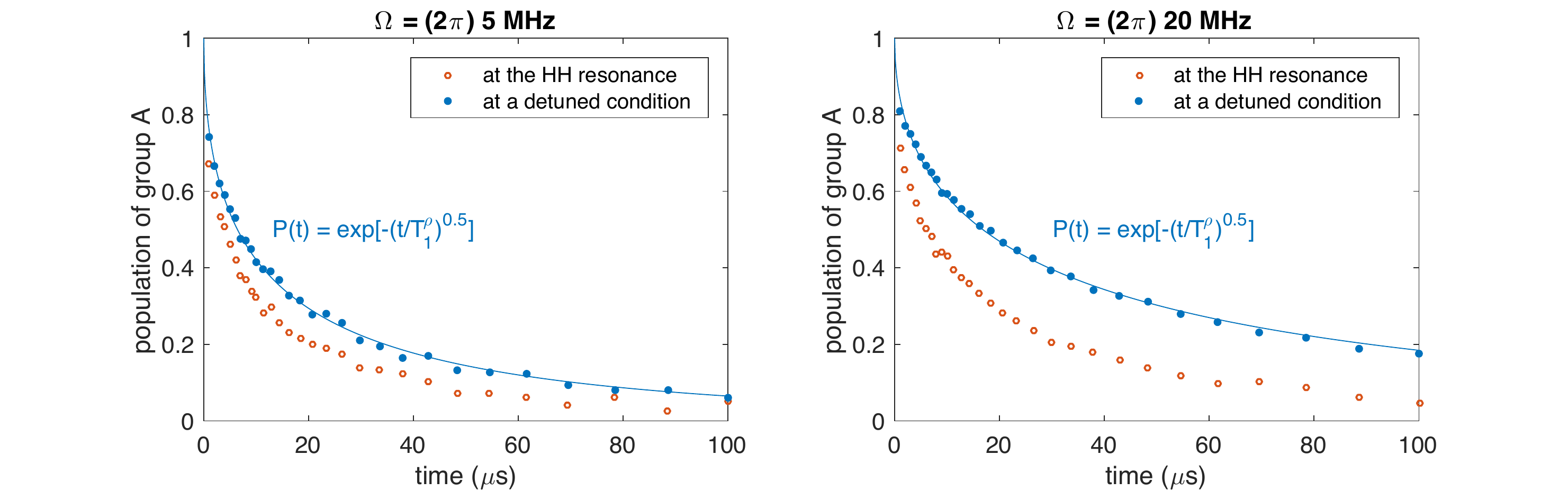}
\end{center}
\caption{\textbf{Unnormalized experimental Data.} Two data sets with different common Rabi frequencies of $\Omega$ = (2$\pi$) [5, 20] MHz are presented at the Hartmann-Hahn resonance (red) and at the far-detuned case (blue). For the detuned signal, a stretched exponential of power 0.5 is fitted to the data. }
\label{fig:rawdata}
\end{figure}

In our Hartman-Hahn experiments, the spin dynamics are governed by both coherent cross-relaxation and incoherent depolarization.
These two effects have qualitatively different dependence on the driving strength and can be clearly distinguished in our observations.
To perform a detailed analysis of the results presented in the main text, we focus on the coherent dynamics by normalizing our data at the Hartman-Hahn resonance $\Omega_A = \Omega_B$ via a sufficiently detuned case $|\Omega_A - \Omega_B| \gg nJ_0$, at which the spin relaxations are dominated by incoherent dynamics (Fig.~\ref{fig:rawdata}, blue line).
Such normalization can be justified only if the two effects are independent and multiplicative.
This is the case if the incoherent dynamics are induced by an independent Markovian noise, which results in an exponential and multiplicative factor $e^{-\gamma t}$.  
In our experiment, however, we observe a stretched exponential $e^{-\sqrt{t/T}}$ decay profile from incoherent dynamics (Fig.~2D in the main text).
Below, we explain why such incoherent decays are still factorizable.

Our incoherent dynamics can be modeled as follows. (See Ref.~\cite{SinkPaper} for more details).
Each spin at site $i$ undergoes incoherent depolarization at rate $\gamma_i$.
This rate $\gamma_i$ is determined by the microscopic local environment of the spin and follows a random distribution $\rho(\gamma; T_1^\rho)$, such that the ensemble averaged polarization decays as a stretched exponential
\begin{align}
    e^{-\sqrt{t/T_1^\rho}} =\int_0^{\infty} \rho(\gamma; T_1^\rho) e^{-\gamma t} d\gamma.
\end{align}
The analytical expression as well as the microscopic origin of the distribution $\rho(\gamma; T_1^\rho)$ are presented in Ref.~\cite{SinkPaper}.
At the Hartman-Hahn condition, both the incoherent process and the coherent cross-relaxation lead to depolarization (see Fig.~\ref{fig:rawdata}).
Hence, at time $t$, the rate of depolarization for spin-$i$ is given by
$\dot{p}_i (t) = - [\gamma_i   + f_i (t) ] p_i (t)$, 
where $f_i (t)$ is the rate of cross-relaxation (which generally depends on the state of other spins). This cross-relaxation, once averaged over an ensemble, leads to a power-law decay as derived in the previous section. The differential equation for the polarization is exactly solvable with the solution $ p_i (t) = e^{-\gamma_i t} e^{\int_{0}^t f_i (t')dt' }$, where one finds a multiplicative exponential factor $e^{-\gamma_i t}$.
Crucially, this effect is still factorizable, even after ensemble averaging:
\begin{align}
    \left\langle p_i(t) \right\rangle_\textrm{ensemble} = \int_0^\infty \rho(\gamma; T_1^\rho)  e^{-\gamma_i t} d\gamma  \left\langle e^{\int_0^t f_i (t') dt'} \right\rangle_\textrm{ensemble} \propto e^{-\sqrt{t/T_1^\rho}} \cdot t^{-\eta},
\end{align} 
where $\eta$ is the disorder dependent exponent derived in the main text.
Physically, this factorization arises because the microscopic environment for each spin, which determines coherent as well as incoherent dynamics, is random and independent. For this reason, in the experiment, we normalize the polarization decay at the Hartmann-Hahn resonance (Fig.~\ref{fig:rawdata}, red line) by the incoherent decay at the far-detuned case (Fig.~\ref{fig:rawdata}, blue line).

\subsection{Dependence of Thermalization Dynamics on Spin Bath Polarization}
\begin{figure}[h!]
\begin{center}
\includegraphics[width=0.9\textwidth]{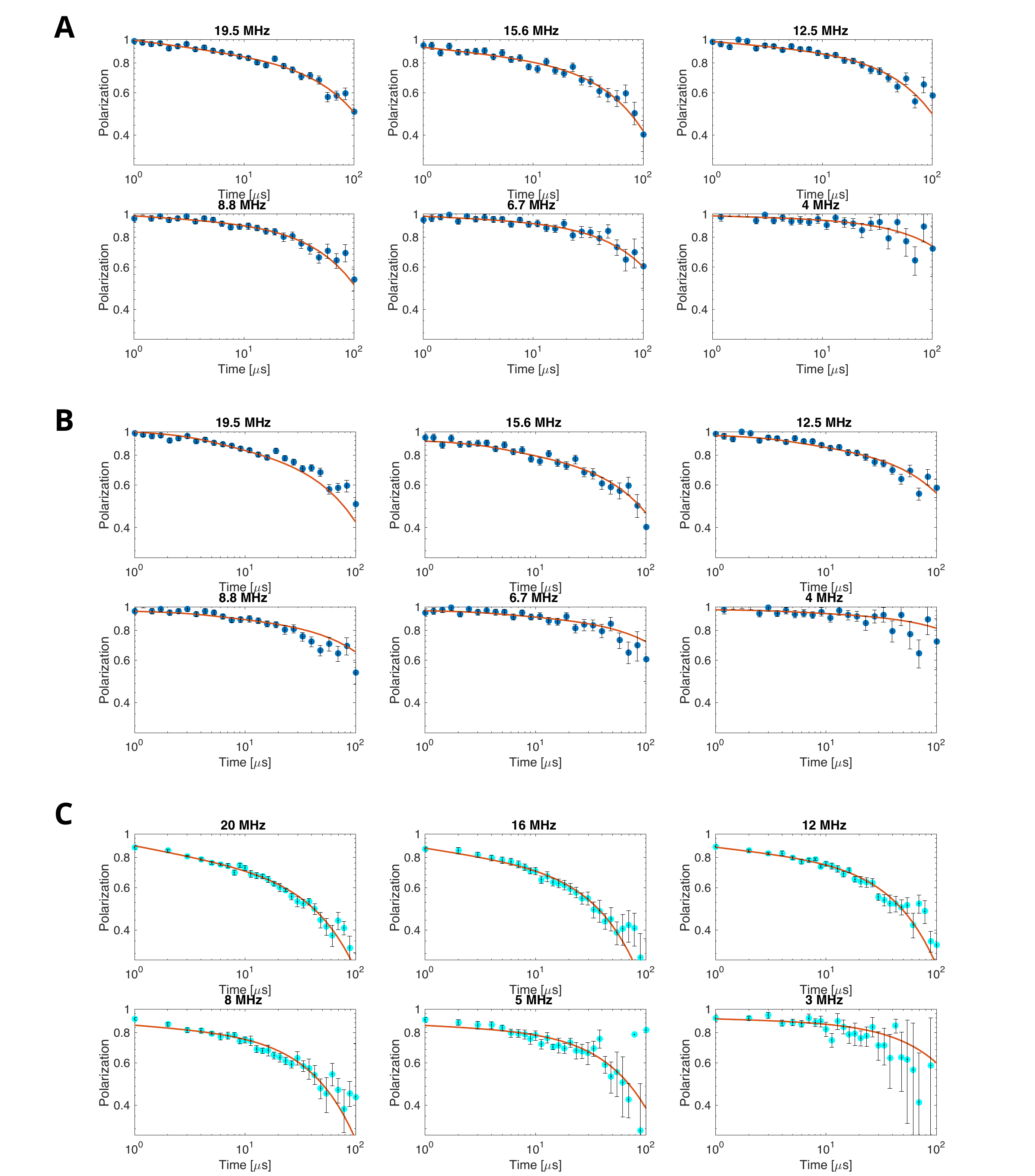}
\end{center}
\caption{\textbf{Polarization decay of a NV ensemble under Hartmann-Hahn conditions.} An initially polarized group A spin ensemble interacts with \textbf{(A,B)} unpolarized and \textbf{(C)} fully polarized group B. Solid lines are theoretical fits based upon a time-dependent disorder model with extracted parameters ($W_d, \tau_d$) via a Monte-Carlo optimization. In \textbf{(A,C)}, the spectral diffusion time $\tau_d$ is dependent on the Rabi frequency, while in \textbf{(B)} $\tau_d$ is independent of the applied Rabi frequency. The dynamical disorder strength $W_d$ is a $\Omega$-independent, global variable in all three cases.}
\label{fig:HHfits}
\end{figure}

\begin{figure}[h!]
\begin{center}
\includegraphics[width=1\textwidth]{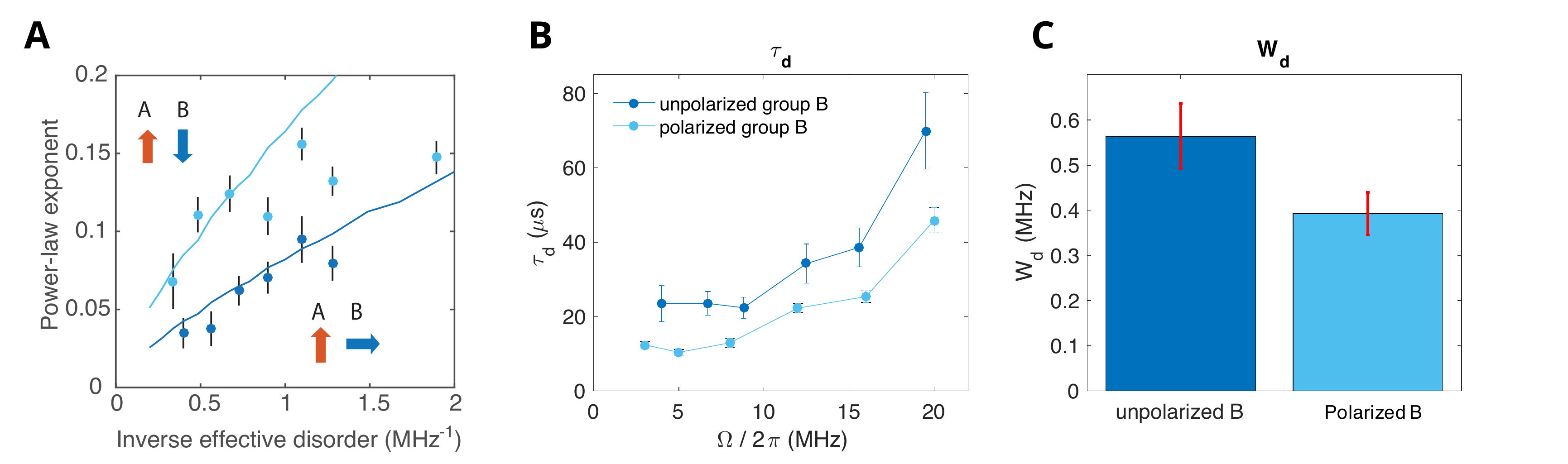}
\end{center}
\caption{ \textbf{Fitted parameters of the time-dependent disorder model extracted from a Monte-Carlo optimization.} \textbf{(A)} Exponents of the power-law decay of group A polarization with oppositely polarized (light blue) and unpolarized (dark blue) group B as a function of inverse effective disorder. Solid lines correspond to numerical simulation results. \textbf{(B)} The extracted $\tau_d$ as a function of Rabi frequency. Light and dark blue point corresponds to fully oppositely polarized and unpolarized group B spin states, respectively. \textbf{(C)} The extracted dynamical disorder $W_d$. All errorbars are evaluated from the standard deviation of the optimized parameter after running 10 independent Monte-Carlo runs.}
\label{fig:HHfitparam}
\end{figure}

In Fig.~\ref{fig:HHfits}, the theory prediction from Eq.~(\ref{eqn:tddisorder_2_surv_prob}) is compared with experimental data for various Rabi frequencies and two different initial polarizations of group B spins. The functional profiles of the decay are consistent with power laws for over a decade, followed by accelerated, though still sub-diffusive relaxation at late times.
In the power-law regime, we find that the power-law decay exponents depend on the initial polarization of group B spins (Fig.~\ref{fig:HHfitparam}A). 
This is consistent with our theory; for single-particle dynamics, we expect that the power-law exponent scales as $\sim nJ/W_\textrm{eff}$, where $n$ is the density of oppositely polarized spins. Indeed, when group B is initially unpolarized, the exponents are decreased by a factor of two compared to the fully polarized case, consistent with our theory at a quantitative level.

To characterize the late-time acceleration of the polarization decay, we use the time-dependent model where the pair-resonance counting criteria are modified as discussed in the previous section. By fitting the experimental data to our model using a Monte-Carlo (MC) optimization, we extract the parameters of the dynamical disorder strength $W_d$ and spectral diffusion time $\tau_d = 1/\Gamma$. 
Here, we assume $W_d$ as a global fit parameter which is independent from $\Omega$; this is because we expect $W_d$ to be predominantly determined by the mean-field interaction strength. 
In contrast, $\tau_d$ may in principle be dependent on $\Omega$ since the fluctuations of the Ising mean-field potential depend on the thermalization speed and hence also on the effective disorder strength tuned by $\Omega$. 
To this end, we performed two independent MC optimizations where we (i) treat $W_d$ as global ($\Omega$-\textit{independent}) and $\tau_d$ as local ($\Omega$-\textit{dependent}) variables (Fig.~\ref{fig:HHfits}A), or instead (ii) fix both parameters as global variables (Fig.~\ref{fig:HHfits}B).
For the static effective disorder, we use the theory-predicted values $W_s \sim W^2/\Omega$ (as described previously).
Naturally, owing to the larger number of fit parameters, a global $W_d$ with local $\tau_d$ variation (Fig.~\ref{fig:HHfits}A) shows better agreement than a global $W_d$ together with with global $\tau_d$ (Fig.~\ref{fig:HHfits}B). In the latter case, extracted fit parameters $W_d$ and $\tau_d$ are $(2\pi) ~46 \pm 14$ kHz and $43 \pm 9~\mu s$, respectively. 

In the former case (Fig.~\ref{fig:HHfits}A), the extracted dynamical disorder $W_d \sim (2\pi)~0.5$ MHz is consistent with the expected strength of the Ising interaction, suggesting that spin-spin interactions play an important role for the time-dependent disorder.
Furthermore, $\tau_d$ is also consistent with the observed NV depolarization timescale, including contributions from both coherent cross-relaxation and incoherent spin depolarization. We note that in the fully polarized case the extracted values for $\tau_d$ are smaller than those in the unpolarized case (Fig.~\ref{fig:HHfitparam}B). We speculate that this could be due to faster coherent spin-exchange dynamics in the former case, giving rise to a faster fluctuation in $\delta^I$, responsible for the accelerated thermalization dynamics at late times.

\clearpage
\bibliography{biblatex_supp}